\documentclass[sigconf]{acmart}

\AtBeginDocument{%
  }

\acmYear{2025}\copyrightyear{2025}
\setcopyright{cc}
\setcctype[4.0]{by}
\acmConference[SEC '25]{The Tenth ACM/IEEE Symposium on Edge Computing}{December 3--6, 2025}{Arlington, VA, USA}
\acmBooktitle{The Tenth ACM/IEEE Symposium on Edge Computing (SEC '25), December 3--6, 2025, Arlington, VA, USA}
\acmDOI{10.1145/3769102.3770629}
\acmISBN{979-8-4007-2238-7/25/12}

\usepackage{algorithm}

\usepackage{algpseudocode}

\usepackage{amsmath}

\begin{document}

\title{\emph{Bayes-Split-Edge}: Bayesian Optimization for Constrained Collaborative Inference in Wireless Edge Systems}

\author{Fatemeh Zahra Safaeipour}
\email{fzsafaei@ku.edu}
\affiliation{%
  \institution{University of Kansas}
      \city{Lawrence}
      \state{Kansas}
      \country{USA}
}
\author{Jacob Chakareski}
\email{jacobcha@njit.edu}
\affiliation{%
  \institution{New Jersey Institute of Technology}
    \city{Newark}
    \state{New Jersey}
    \country{USA}
}

\author{Morteza Hashemi}
\email{mhashemi@ku.edu}
\affiliation{%
  \institution{University of Kansas}
    \city{Lawrence}
    \state{Kansas}
    \country{USA}
}

\renewcommand{\shortauthors}{Safaeipour et al.}

\begin{abstract}
Mobile edge devices (e.g., AR/VR headsets) typically need to complete timely inference tasks while operating with limited on-board computing and energy resources. In this paper, we investigate the problem of collaborative inference in wireless edge networks, where energy-constrained edge devices aim to complete inference tasks within given deadlines. These tasks are carried out using neural networks, and the edge device seeks to optimize inference performance under energy and delay constraints. The inference process can be split between the edge device and an edge server, thereby achieving collaborative inference over wireless networks.
We formulate an inference utility optimization problem subject to energy and delay constraints, and propose a novel solution called Bayes-Split-Edge, which leverages Bayesian optimization for collaborative split inference over wireless edge networks. Our solution jointly optimizes the transmission power and the neural network split point. The Bayes-Split-Edge framework incorporates a novel hybrid acquisition function that balances inference task utility, sample efficiency, and constraint violation penalties.
We evaluate our approach using the VGG19 model on the ImageNet-Mini dataset, and Resnet101 on Tiny-ImageNet, and real-world mMobile wireless channel datasets. Numerical results demonstrate that Bayes-Split-Edge achieves up to 2.4× reduction in evaluation cost compared to standard Bayesian optimization and achieves near-linear convergence. It also outperforms several baselines, including CMA-ES, DIRECT, exhaustive search, and Proximal Policy Optimization (PPO), while matching exhaustive search performance under tight constraints.
These results confirm that the proposed framework provides a sample-efficient solution requiring maximum 20 function evaluations and constraint-aware optimization for wireless split inference in edge computing systems.
\end{abstract}


\begin{CCSXML}
<ccs2012>
   <concept>
       <concept_id>10003033.10003106.10003113</concept_id>
       <concept_desc>Networks~Mobile networks</concept_desc>
       <concept_significance>300</concept_significance>
       </concept>
   <concept>
       <concept_id>10010147.10010178.10010219</concept_id>
       <concept_desc>Computing methodologies~Distributed artificial intelligence</concept_desc>
       <concept_significance>500</concept_significance>
       </concept>
   <concept>
       <concept_id>10002950.10003714.10003716.10011141</concept_id>
       <concept_desc>Mathematics of computing~Mixed discrete-continuous optimization</concept_desc>
       <concept_significance>300</concept_significance>
       </concept>
 </ccs2012>
\end{CCSXML}

\ccsdesc[300]{Networks~Mobile networks}
\ccsdesc[500]{Computing methodologies~Distributed artificial intelligence}
\ccsdesc[300]{Mathematics of computing~Mixed discrete-continuous optimization}

\ccsdesc[500]{Computing methodologies~Distributed algorithms}

\keywords{Collaborative Inference, Split Learning, Bayesian Optimization, Wireless Edge Computing, Constrained Optimization, Mobile Systems, Resource Allocation, Neural Networks, VR / AR / MR, Cloud-Edge-Device Continuum, Metaverse.}

%\received{22 June 2025}

\maketitle

\section{Introduction}
The rapid expansion of emerging power and computation limited mobile edge devices, VR/AR/MR headsets, and smart glasses, equipped with multiple cameras and other sensing modalities, and novel applications in which they are integrated, introduces challenges in efficient data processing and transmission \cite{8234686,ChakareskiKRB:21,GuptaCP:20,BadnavaCH:24a,ChakareskiK:23}.
%power-limited devices, such as mobile phones and various edge devices, has introduced challenges in efficient data processing and transmission \cite{8234686}. 
Such devices typically generate substantial data volumes that require timely and efficient processing and distribution, in many cases over the Cloud-Edge-Device continuum. Yet, the devices' limited power resources and unreliable wireless channels introduce several technical challenges \cite{mao2017surveymobileedgecomputing}. In particular, power constraints hinder the ability of these devices to either process data locally or transmit all raw data to an edge server for computation. Furthermore, computation tasks must be completed within a specific deadline, and dynamic channel conditions introduce variability that must be considered.
\begin{figure}[ht]
    \centering
    \includegraphics[width=0.48\textwidth]{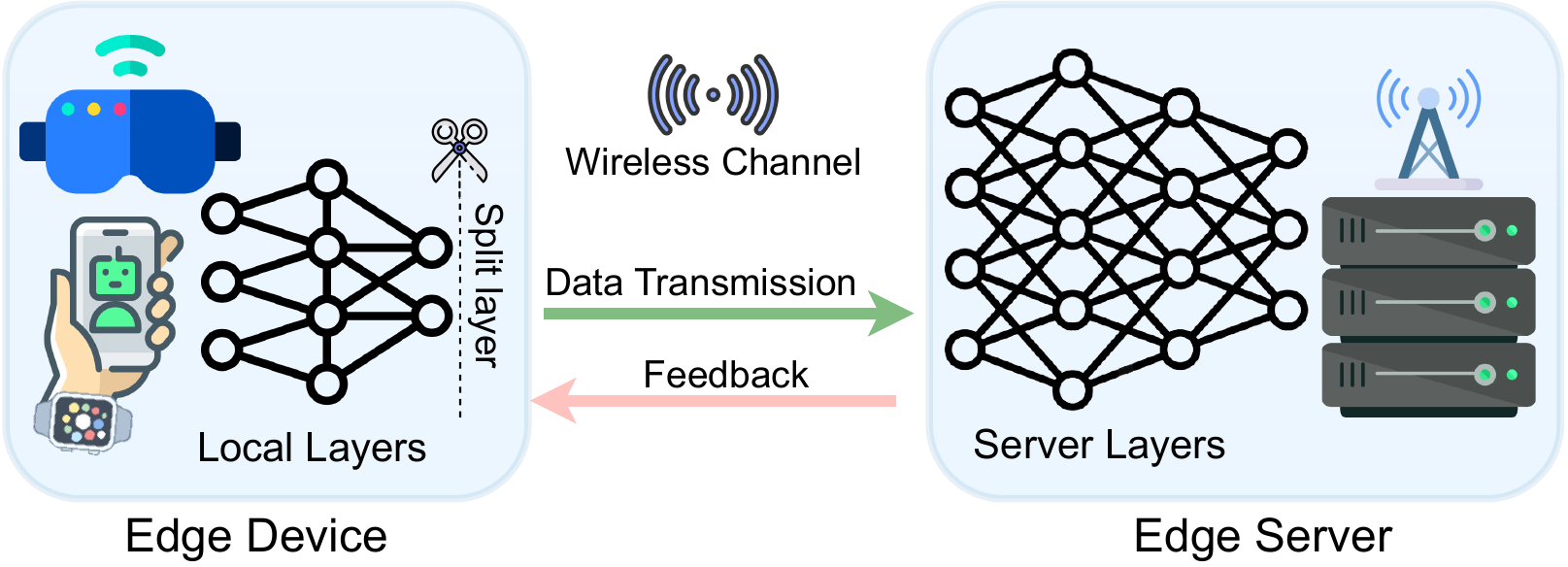} 
    \caption{System overview of wireless split learning. The edge device (e.g., AR headset, mobile phone, or wearable) performs initial neural network layers locally, while the remaining layers are offloaded to the edge server. Intermediate features are transmitted over a wireless channel. Feedback on network conditions is used to adapt the split layer dynamically to optimize performance under resource constraints.}
    \label{fig:system_model}
\end{figure}

The core challenge is to balance power consumption between local computation and data transmission to the edge server to ensure that the power and delay constraints are met~\cite{10620728}. Current approaches either process data entirely at the edge server~\cite{10040976}, incurring high transmission costs, or rely on local computation, which quickly exhausts the device's power. Split learning \cite{10038613} offers a middle ground, enabling devices to process initial computations locally, reducing data transmission to a manageable volume while offloading the remaining computations to the edge server. 

Recently, several research studies have focused on split learning in different scenarios. For example, the authors in \cite{thapa2020splitfed} explore the integration of split learning with federated learning \cite{ni2025pfedwn} to enhance privacy and scalability. Avasalcai et al. \cite{8812207} and Yang et al.\cite{lin2024adaptsfladaptivesplitfederated} extended split learning to mobile and adaptive networks, but lacked real-time power adaptation. The research work in~\cite{9224971} and \cite{5963233}
address energy efficiency in static settings, while the authors in \cite{9507488} focused on power-saving in IoT networks. %Unlike these static frameworks, BROWSER dynamically balances power and delay, validated in fluctuating wireless environments, demonstrating its ability to adjust split points and achieve significant power savings and reduced latency.
%\noindent \textbf{Energy-Efficient Wireless Networks.} 
In parallel, extensive amounts of work (see, for example, \cite{7890237, 7448820, ghazikor2024channel}) have focused on transmission efficiency without addressing computation offloading.  
% BROWSER achieves balanced power-delay optimization through dynamic split selection, validated in simulations, and demonstrates reduced transmission energy compared to communication-focused approaches.
%\noindent \textbf{Adaptive Split Learning.}
Adaptive split learning methods, such as \cite{9355679, 9359147},
enhance flexibility by adjusting split points, but do not incorporate power constraints.% BROWSER addresses this gap, reducing power consumption through real-time split adjustments and delivering average power savings of 18% compared to conventional SL.
%Privacy-preserving SL \cite{khan2023more, thapa2021advancements}  has been applied in data-sensitive contexts, prioritizing privacy over power efficiency and latency. 

Within this context, existing research works on split learning techniques focus primarily on computational or transmission efficiency, often overlooking the stringent power and delay constraints that resource-limited devices in wireless networks face.
% However, optimizing split learning for real-time applications in resource-constrained and wireless settings introduces additional challenges and complexities\cite{10314792}.
To address this gap, in this paper, we develop an efficient split learning solution for wireless devices with power and delay constraints. In particular, our system model shown in Figure~\ref{fig:system_model} consists of a energy-limited mobile device that processes data (i.e., computation task) locally up to a split layer \( l \) to reduce transmission load and offloads the remaining computations to an edge server. To determine the optimal split layer $l$ and allocate transmission power while meeting energy and delay constraints, we formulate an optimization problem aimed at maximizing computational utility. To solve the formulated problem, we develop a framework using Bayesian Optimization (BO)\cite{papenmeier2024bouncereliablehighdimensionalbayesian}. 
% which has emerged as an effective tool for managing high-dimensional environments . 
BO uses probabilistic models to efficiently explore the search space, making it well-suited for scenarios that require adaptive decision-making under uncertainty %snoek2012practical, shahriari2015taking, 
\cite{frazier2018tutorialbayesianoptimization, lu2022surrogatemodelingbayesianoptimization,10234224}. Beyond empirical performance, BO based on Gaussian Processes (GPs) offers provable guarantees. The GP‐UCB algorithm achieves sublinear cumulative regret by balancing exploration and exploitation via information gain~\cite{Srinivas_2012}. In practice, this ensures rapid convergence even when the objective is complex or stochastic. Together, BO’s demonstrated efficiency and theoretical foundations make it a suitable choice for black‐box optimization in noisy, resource-constrained wireless environments.

In this paper, we leverage BO to select the optimal split points for computation offloading to the edge server, and allocates optimal transmit power so that the power and delay constraints are satisfied. 
To capture the constraints, we propose a novel hybrid acquisition function for the BO framework and establish an integrated  method called Bayes-Split-Edge. 
We evaluate the proposed solution using the VGG19 model on the ImageNet-Mini dataset, along with the real-world mMobile mobility dataset to emulate realistic wireless channel conditions. Our numerical results show that Bayes-Split-Edge  outperforms several other baselines, including CMA-ES, DIRECT,  Proximal Policy Optimization (PPO) baselines. Our results show that the proposed method achieves near-optimal computation accuracy performance while providing a fast convergence rate and meeting the power and delay constraints.   In summary, our main contributions are as follows:

\begin{itemize}
\item \emph{Bayesian Optimization Framework for Split Inference.} We introduce Bayes-Split-Edge framework that jointly selects the optimal neural network split point and transmission power to maximize inference performance under strict energy and delay constraints in wireless edge networks.

\item \emph{Hybrid Acquisition Function for Constraint-Aware Optimization.} We propose a new hybrid acquisition function that incorporates expected improvement, uncertainty-based exploration, gradient-based stability, and soft penalties for constraint violations. The proposed acquisition function improves the sample efficiency and performance compared with a standard acquisition function.   

\item \emph{Performance Evaluation on a Realistic Setup.} We present extensive evaluation results using the VGG19 model on ImageNet-Mini with real-world mMobile wireless traces. The results verify that the proposed method achieves faster convergence and higher accuracy compared to several baselines, including standard BO, CMA-ES, PPO, and exhaustive search.
% \item We propose an efficient split learning method, named Bayes-Split-Edge, that is tailored for power-constrained devices in wireless networks. The proposed framework dynamically adjusts the split point and power allocation for computation and communication tasks to meet power and delay constraints. Our proposed solution optimally allocates resources using Bayesian Optimization (BO) with constraint handling and gradient updates.

% \item We introduce a Bayesian Optimization (BO) framework based on a Gaussian process that incorporates variability of the wireless channel, power limitations, and task deadlines. This method consists of a refined acquisition function that uses penalty terms and gradient-based updates to improve performance. Using BO's probabilistic and model-driven features, our solution dynamically adjusts the split points and power allocation. 

% \item Through extensive simulations, we demonstrate that Split-Edge achieves faster convergence, higher resource allocation efficiency, and optimized computation accuracy compared to traditional approaches while satisfying power and delay requirements. 
\end{itemize}
\noindent 
The rest of this paper is organized as follows. Section \ref{sec:related-work} presents a summary of related works. In Section \ref{sec:system-model}, we present our proposed system model and problem formulation. Section \ref{sec:solution} includes our developed solution, followed by numerical results in Section \ref{sec:results}. Finally, Section \ref{sec:conclusion} concludes the paper.

\section{Related Work}
\label{sec:related-work}

\noindent\textbf{Constraint-aware Split Inference.}
The authors in \cite{10279444} present an online algorithm based on Lyapunov stochastic optimization to choose CNN split points and uplink rates to minimize device energy consumption, while satisfying end‐to‐end delay constraints in a single‐user edge inference setting. However, by assuming error‐free transmission at supported rates, the framework does not incorporate the impacts of wireless channel impairments that lead to degraded inference accuracy.
Furthermore, the study in \cite{10478867} formulates a mixed‐integer nonlinear program to minimize total energy consumption across multiple sensor devices under a latency constraint in a multi‐user wireless sensing system. They propose LOP algorithm, which combines a deep reinforcement learning model for optimal split‐point selection with convex optimization for resource allocation, and demonstrate near‐optimal energy efficiency and reduced computation delay compared to full local or full offload schemes. However, their framework assumes error‐free transmission and does not model how channel impairments or expanding intermediate feature sizes can degrade inference accuracy. Additionally, the LOP policy network requires on the order of two to three thousand training epochs to converge, posing challenges for timely adaptation in dynamic environments.

\noindent\textbf{Unconstrained Split Inference Optimization.}
In another line of work, the authors in \cite{9685179} introduce SI-NR, a split‐inference framework that trains deep neural networks with dropout to tolerate up to $60\%$ packet loss and thus eliminate retransmission delays in lossy IoT networks. By emulating packet drops during training, SI-NR maintains high prediction accuracy without any retransmissions, ensuring low-latency inference over unreliable links. However, the approach does not account for edge‐device limitations, such as on‐device compute capacity, memory footprint, and energy consumption, which are essential for practical deployment on resource‐constrained IoT hardware.

The paper \cite{10107635} proposes a two‐stage split‐inference framework that uses an offline, exhaustive search to determine U‐shaped DNN partitions (edge‐side and server‐side) under memory and energy constraints.  Then adaptively selects the split point in real time based on instantaneous channel gains to minimize average latency while preserving privacy. However, the approach does not account for potential inference accuracy degradation when intermediate data are transmitted over poor channels, and its offline exhaustive search yields a static partition that cannot evolve or improve over time as network conditions or model behavior change.
The paper \cite{9933908} proposes an adaptive edge inference framework that integrates multi‐exit DNNs with model partitioning to serve multiple mobile inference streams. Under the assumption of known task arrivals, an offline dynamic programming algorithm selects exit and partition points to optimize the tradeoff between processing latency and inference accuracy. An online learning‐based algorithm, enhanced by prioritized experience replay and historical initialization, dynamically adjusts these points in real time. Nonetheless, the framework does not consider wireless channel variability or edge‐node resource constraints, and it requires approximately $1,000–2,000$ training epochs to converge, which may be impractical in rapidly changing environments.
% ULO \cite{machines10080629} is a lightweight version of a heavyweight detector, built to run on small underwater devices with limited battery and compute power. It uses clever design tricks to slim down a larger model so it can squeeze into these tight environments. However, because its slimmed-down shape is fixed at design time, it can’t shrink any further if a particular device has even fewer resources than anticipated. In other words, if ULO’s pared-down version still doesn’t fit on some gadget, it has no built-in way to scale back even more.
In ISCC \cite{11016266}, the authors propose a framework that jointly optimizes split inference, model pruning, and feature quantization to minimize edge‐device energy under accuracy and latency constraints. They derive an offline inference‐accuracy model and solve a nonconvex resource‐allocation problem—over pruning ratio, split layer, quantization level, sensing power, and transmit power—by enumerating split/quantization pairs and running alternating KKT‐based and golden‐section updates until convergence. Simulations show up to $40\%$ energy savings in low‐latency scenarios; however, the offline accuracy approximation and static allocations prevent adaptation to runtime channel or device changes.

%\noindent\textbf{Cloud Split Inference.}
Similar studies \cite{mudvari2024adaptive, 8871124,sac, 10529950, 10646420, 10621218, 9384272} have explored splitting computations among device, edge server, and cloud. However, because cloud communication incurs excessive latency and cannot meet real-time requirements, we do not consider cloud-based architectures.

\noindent\textbf{Sample-efficient Optimization.}
Relative to prior studies, we leverage Bayesian Optimization (BO) over reinforcement‐learning (RL) methods like DDPG primarily, for sample efficiency. To demonstrate the sample-efficiency of BO, for example, in \cite{9414155}, BO identified Pareto‐optimal configurations in a cellular network with only 1,012 evaluations, whereas DDPG required over 600,000. In general, RL methods typically require large amounts of training iterations to converge~\cite{9430561}.
% BO has also proven effective in online radio resource management. For example, tuning uplink power control under noisy, black‐box objectives, minimizing performance drops with relatively few evaluations \cite{9430561}. 
On the other hand, low evaluation cost, noise resilience, and adaptability to dynamic constraints, make BO well-suited for wireless split learning, where utility functions depend on varying channel conditions and lack closed‐form expressions. In this paper, we enhance the basic BO by developing a hybrid acquisition function that integrates the inference accuracy, sample efficiency, and constrain violation penalty. 

\section{Motivating Example: Profiling an Inference Model}
Here, we present empirical \emph{profiling} results that reveal the fundamental complexity of optimal layer splitting in collaborative inference over wireless edge networks. Our analysis demonstrates that the optimal split layer is not static but dynamically depends on multiple interdependent factors, making this a challenging multi-objective optimization problem. 
\subsection{Experimental Setup and Methodology.}
We profile the VGG19 deep convolutional neural network, a representative CNN architecture for image classification, across all possible split points. Our experimental framework employs real-world mobile channel traces (mMobile \cite{mmobile}) to emulate time-varying wireless conditions between edge devices and servers. The channel gain fluctuations directly impact achievable data rates and, consequently, transmission delays for intermediate activations.
The results presented here correspond to a single transmit power level. In practice, as we have considered, edge devices can dynamically adjust their transmit power, introducing an additional optimization dimension that further complicates the splitting decision space.
%We consider the VGG19 model that is a deep convolutional neural network (CNN) model for image classification tasks. We use real-world mMobile datasets to emulate wireless channel condition between edge device and server. Depending on the channel gain, the achieved data rate varies over time, which in turn, impacts transmission delays. 

\subsection{Empirical Analysis}
\begin{figure}[t]
    \centering
    \includegraphics[width=.48\textwidth]{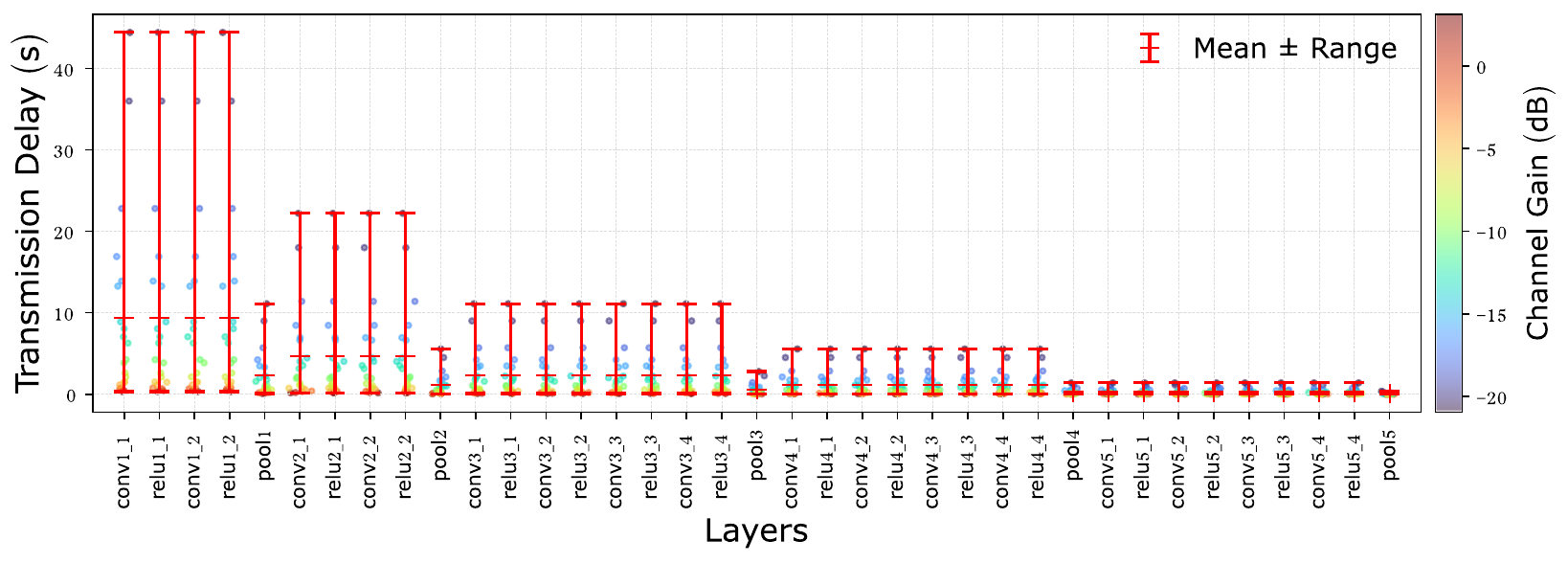} 
    \vspace{-.4cm}
    \caption{Transmission delay across different split layers under varying channel conditions. The red error bars indicate the mean and range (max–min) of delay measurements over multiple frames. Background color represents the corresponding channel gain (in dB).}
    \label{fig:transmission_delay}
\end{figure}

Figure ~\ref{fig:transmission_delay} illustrates the transmission delay characteristics across different split layers under varying channel conditions. The red error bars represent the mean and range (maximum-minimum) of delay measurements across multiple channel realizations. The background color intensity indicates the corresponding channel gain in dB. 
Several critical observations emerge from Fig.~\ref{fig:transmission_delay}. (1) High Variability in Early Layers: Split layers in the initial convolutional layers (conv1-1 through conv2-2) exhibit extreme transmission delay variability, with ranges spanning up to 45 seconds under poor channel conditions. (2) Behavior Over Split Layers: As we progress deeper into the network, transmission delays becomes smaller due to the substantial dimensionality reduction achieved by pooling operations and the transition to fully connected layers. (3) Architecture-Dependent Behavior: There is no direct linear relationship between layer index and transmission delay. Instead, delays are dictated by the specific architecture characteristics of each layer, including filter dimensions, feature map sizes, and pooling operations. (4) Channel Dependency: The optimal split layer from a transmission delay perspective is highly sensitive to instantaneous channel conditions. Therefore, we need an adaptive splitting strategies.

%To investigate the delay performance, Fig. \ref{fig:transmission_delay} shows the transmission delay across different split layers under varying channel conditions. The red error bars indicate the mean and range (max-min) of delay measurements over multiple frames. Background color represents the corresponding channel gain (in dB). 

\begin{figure}[h]
    \centering
    \includegraphics[width=0.48\textwidth]{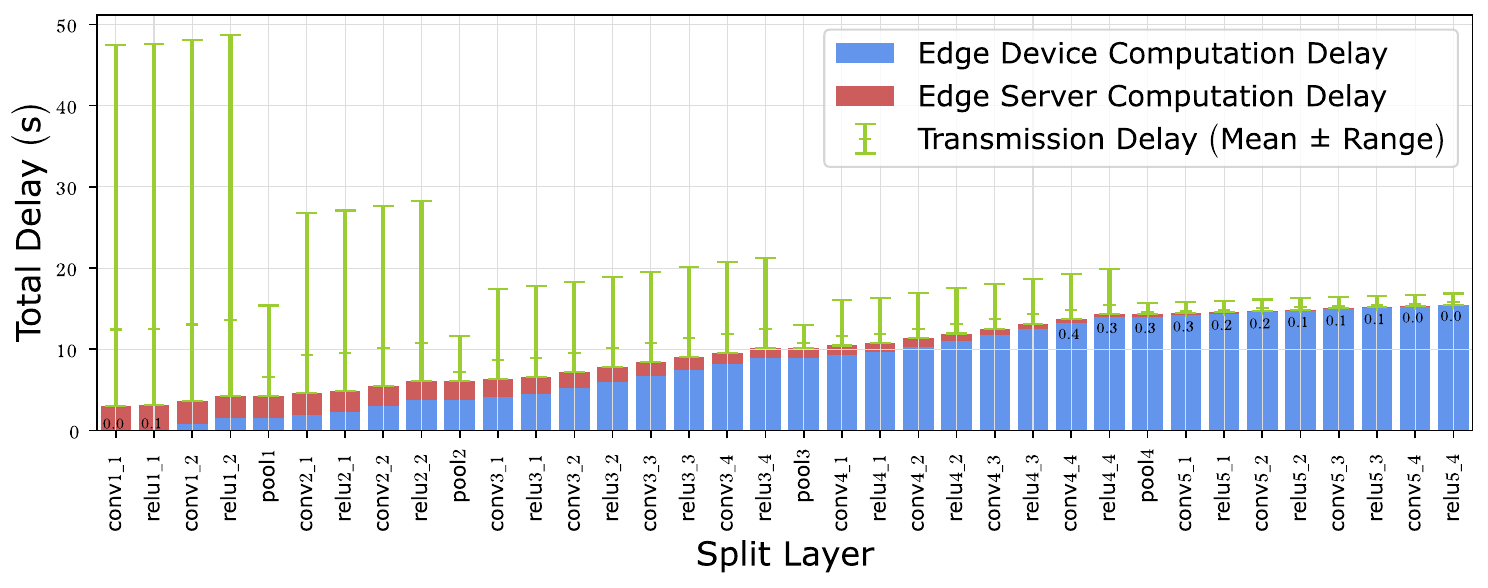} 
    \caption{End-to-end delay breakdown for different split layers in the collaborative inference pipeline. Blue and red bars represent computation delay at the edge device and edge server, respectively, while green error bars indicate the mean and range of transmission delay across multiple channel realizations. We assume negligible server-side transmission delay since the downstream payload (logits/labels) is small compared to the available channel capacity.}
    \label{fig:total_delay}
\end{figure}

Furthermore, Fig.~\ref{fig:total_delay} presents the end-to-end delay breakdown for different split layers in the collaborative inference pipeline. Blue and red bars represent computation delay at the edge device and edge server, respectively, while green error bars indicate the mean and range of transmission delay across multiple channel realizations.
From the results in Fig.~\ref{fig:total_delay}, we observe that: (1) Early splits minimize computation delay but incur prohibitive transmission costs, especially under poor channel conditions. (2) Server-side computation delays are consistently lower than edge-side delays due to superior computational resources, while edge-side computation grows with split depth as more layers are processed locally. This suggests we prefer server computation when this does not lead to high transmission delays (transmission delay is not dominant). (3) The dominant delay component transitions from transmission (early splits) to computation (late splits).

\begin{figure}[h]
    \centering
    \includegraphics[width=0.48\textwidth]{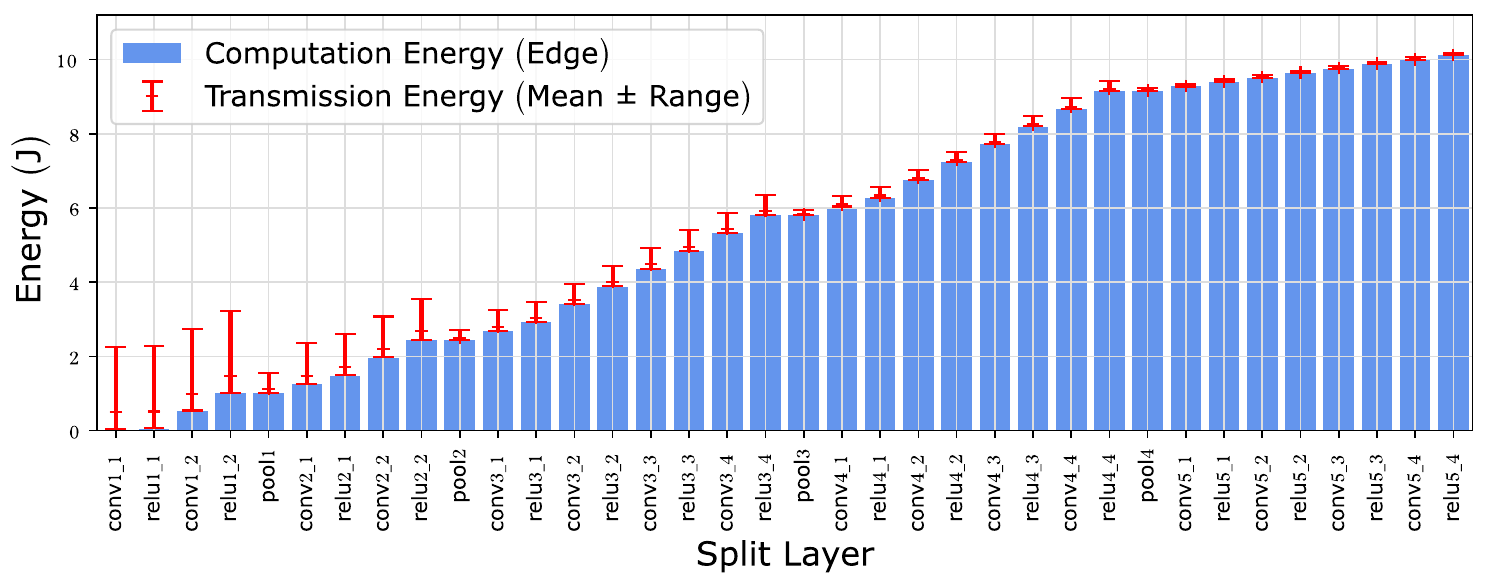} 
    \caption{Energy consumption breakdown across different split layers. Blue bars represent cumulative computation energy on the edge device,and red error bars indicate the mean and range of transmission energy measured over multiple frames. Early splits incur higher transmission energy due to larger activation sizes, while deeper splits increase computation energy as more layers are processed locally.}
    \label{fig:total_energy}
\end{figure}

In terms of energy consumption, Fig. \ref{fig:total_energy} depicts the energy consumption breakdown across different split layers. Blue bars represent cumulative computation energy on the edge device, while red error bars indicate the mean and range of transmission energy measured over multiple frames. Early splits incur higher transmission energy due to larger activation sizes, while deeper splits increase computation energy as more layers are processed locally.
From the results, we observe that finding optimal split point for collaborative inference highly depends on the characteristics of various layers in the neural network as well as underlying channel condition.

\subsection{Problem Complexity}
The profiling results demonstrate that collaborative inference optimization is inherently complex due to several factors: \textbf{(1) Multi-dimensional Optimization Space}: The optimal splitting decision must simultaneously navigate temporal dynamics as channel conditions vary over time, constraints imposed by hardware specific resources of edge device, and power control considerations that add another optimization dimension beyond the single-power scenarios analyzed here. \textbf{(2) Non-convex Objective Landscape}: The empirical results suggest that the optimization landscape is non-convex with multiple local optima. The optimal split point can shift dramatically with small changes in channel conditions or system resources. \textbf{(3) Stochastic System Behavior}: The large error bars in transmission-related metrics highlight the stochastic nature of wireless channels. Any practical solution must account for this uncertainty and converge fast enough to keep up with varying conditions.

\subsection{Implications for Algorithm Design}
Our analysis reveals three algorithmic requirements. First, optimal layers become obsolete within seconds, making traditional approaches requiring more than 50 iterations for convergence fundamentally inadequate because they can not adapt fast enough. Second, algorithms must converge within a fraction of channel coherence time under ultra-high sample efficiency constraints, as each optimization step must extract maximum information from limited observations. Third, function evaluations require actual inference execution with tangible costs, unlike simulation-based optimization.

The fast fading environment creates a non-stationary optimization problem where the objective function changes faster than it can be explored. Therefore, a successful algorithm must quickly identify promising regions, efficiently balance exploration-exploitation, and converge to near-optimal solutions within severely limited evaluation budgets.
These challenges motivate our adaptive Bayesian optimization solution, \emph{Bayes-Split-Edge}, with hybrid acquisition functions presented in subsequent sections.
\section{System Model and Problem Formulation}\label{sec:system-model}

In this section, we present our envisioned system model, followed by the problem formulation. 

\subsection{System Model}

\noindent \textbf{Setup.} 
We consider a wireless edge computing system consisting of a resource-constrained mobile device (MD) and an edge server (S). The MD, such as a UAV, smartphone, or embedded sensor, generates data and executes a portion of a computing task locally (e.g., the early layers of a DNN or signal processing pipeline). The remaining layers are to be completed by the edge server.

Each task \( k \in \{1, \dots, K\} \) requires executing a computation composed of \( L \) sequential layers. The split layer \( l_k \in \{1, \dots, L\} \) defines how many layers are processed on the device. After processing the first \( l_k \) layers locally, the mobile device (MD) generates an intermediate output of size \( D(l_k) \), which is transmitted to the server over a wireless channel. The original input size is \( D_k \), and we assume \( D(l_k) \le D_k \), reflecting the compression typically achieved by early layers that is an implicit effect of the solution algorithm.

The MD consumes energy for both local computation and data transmission, denoted by \( E_{c,k} \) and \( E_{t,k} \), respectively. These values depend on how many layers are executed locally and the size of the resulting output. The total delay for task \( k \) consists of three components: the local computation delay \( \tau_{c,k}^{\text{MD}} \), the transmission delay \( \tau_{t,k} \), and the server-side computation delay \( \tau_{c,k}^{\text{S}} \). We assume the server’s transmission delay is negligible or zero, either because the transmitted result (e.g., labels or lightweight outputs) is small, or because the goal was to offload data efficiently.

\noindent \textbf{Communication Model.}  
The mobile device transmits the intermediate output \( D(l_k) \) to the edge server over a wireless uplink channel. This channel experiences realistic impairments such as fading and noise, with its quality characterized by the measured channel gain \( h_k \) for task \( k \).

Given the transmission power \( P_{t,k} \), the achievable data rate is:
\begin{equation}
R_k = B \log_2 \left( 1 + \frac{P_{t,k} |h_k|^2}{N_0 B} \right),
\end{equation}
where \( B \) is the channel bandwidth and \( N_0 \) is the noise power spectral density. The transmission rate \( R_k \) varies across tasks based on the selected power level and the corresponding channel conditions. Therefore, the resulting transmission delay is given by:
\begin{equation}
\tau_{t,k} = \frac{D(l_k)}{R_k}.
\end{equation}
which inherits the stochasticity of the wireless channel as shown in Fig.\ref{fig:transmission_delay}. 

\noindent \textbf{Computation Model.}  
Each task is executed jointly by the mobile device (MD) and the edge server. The MD processes the first \( l_k \) layers locally, while the remaining layers \( l_k+1 \) to \( L \) are handled by the server. The local computation energy on the MD depends on the computational cost of each executed layer.

Let \( \alpha_{k,i} \) represent the computational load (e.g., number of multiply-accumulate (MAC) operations) for layer \( i \) in task \( k \), and let \( f \) be the MD's processing frequency. Based on standard models from prior works~\cite{8234686,mao2017surveymobileedgecomputing}, the energy consumed for local computation is:
\begin{equation}
E_{c,k} = \sum_{i=1}^{l_k} \kappa \alpha_{k,i} f^2,
\end{equation}
where \( \kappa \) is a hardware-specific constant that reflects switching capacitance and voltage scaling effects.

The computation delays at the MD and the server are given by:
\begin{equation}
\tau_{c,k}^{\text{MD}} = \sum_{i=1}^{l_k} \frac{\alpha_{k,i}}{f \cdot \eta}, \quad \tau_{c,k}^{\text{S}} = \sum_{i=l_k+1}^{L} \frac{\alpha_{k,i}}{f' \cdot \eta},
\end{equation}
where \( f' \) is the server's processing frequency, and \( \eta \) is the processor efficiency factor.

\subsection{Problem Formulation}

Our objective is to optimize the performance of the split learning system by jointly selecting the split layer \( l_k \in \{1, 2, \ldots, L\} \) and the transmission power \( P_{t,k} \in [P_{\min}, P_{\max}] \) for each task \( k \). We define a utility function \( U_k(l_k, P_{t,k}) \), which reflects task-specific performance, such as classification accuracy, depending on the application. Therefore, we formulate the following optimization problem: 
\begin{subequations}
\begin{align}
    \max_{l_k,\,P_{t,k}} \quad & \frac{1}{K} \sum_{k=1}^{K} U_k(l_k, P_{t,k}) \label{eq:obj} \\
    \text{subject to} \quad 
    & E_{c,k} + E_{t,k} \leq E_{\max}, \label{eq:const_energy} \\
    & \tau_{c,k}^{\text{MD}} + \tau_{t,k} + \tau_{c,k}^{\text{S}} \leq \tau_{\max}, \label{eq:const_delay} \\
    & l_k \in \{1, 2, \ldots, L\}, \label{eq:const_layer} \\
    & P_{t,k} \in [P_{\min}, P_{\max}]. \label{eq:const_power}
\end{align}
\end{subequations}

Here, \( E_{c,k} \) and \( E_{t,k} \) represent the energy used for local computation and wireless transmission, respectively. As in Fig.~\ref{fig:total_energy}, the total energy is dependent on the wireless channel conditions. The delay constraint includes the device's computation time \( \tau_{c,k}^{\text{MD}} \), the uplink transmission time \( \tau_{t,k} \), and the server-side processing time \( \tau_{c,k}^{\text{S}} \), demonstrated in Fig.\ref{fig:total_delay}.
Since the utility function is treated as a black box without a closed-form expression (e.g., classification error), and may be non-smooth or non-convex, gradient-based optimization methods are not directly applicable. We therefore propose a modified Bayesian optimization approach to develop a sample-efficient solution.

\section{Proposed Solution} \label{sec:solution}

We address the problem of jointly optimizing the transmission power \( P_{t,k} \in [P_{\min}, P_{\max}] \) and the split layer \( l_k \in \{1, 2, \dots, L\} \) for each task \( k \) in a wireless split computing system, under energy and delay constraints. The goal is to maximize a task-specific utility function \( U_k(P_{t,k}, l_k) \), such as inference accuracy. This utility is treated as a black-box function: it may be non-convex, non-smooth, and expensive to evaluate. In contrast, the energy and delay constraints are modeled as known, deterministic functions based on analytical expressions. We formulate this as a black-box constrained optimization problem and apply Bayesian Optimization (BO) to sequentially explore the decision space using a surrogate model.

% \subsection{Problem Setting and Assumptions}
\subsection{Gaussian Process Modeling and Iterative Optimization}
Before presenting our solution, we make the following assumptions:  
(1) The utility function \( U_k(P_{t,k}, l_k) \) is deterministic but unknown and can only be evaluated at specific input points.  
(2) The energy \( E_k(P_{t,k}, l_k) \) and delay \( \tau_k(P_{t,k}, l_k) \) constraints are deterministic and derived from known analytical models.  
(3) All tasks share a common utility landscape, owing to similar workloads and channel conditions.  
(4) The inputs are normalized: transmission power \( P_t \in [0,1] \), and the split layer \( l \in [0,1] \) is treated as a continuous variable during optimization and discretized during evaluation.

We model the scalar utility function \( U(\mathbf{a}) \) using a Gaussian Process (GP) surrogate, where the input \( \mathbf{a} = [\tilde{P}_t, \tilde{l}] \in [0,1]^2 \) represents the normalized transmission power and split layer index. Normalization ensures that both input dimensions contribute comparably to kernel distance, which is a standard practice in GP modeling to avoid bias in length-scale estimation.

Since the split layer \( l \in \{1, 2, \dots, L\} \) is discrete by nature, we relax it to a continuous variable in the interval \( [0,1] \) during training. This relaxation enables smooth interpolation across candidate split layers and supports gradient-based acquisition optimization. At evaluation time, the continuous value is rounded to the nearest integer value.

To capture uncertainty in the utility landscape, we fit a zero-mean GP with a Matérn 5/2 kernel, without using automatic relevance determination (ARD). The GP is trained on a dataset \( \mathcal{D}_n = \{(\mathbf{a}_i, U(\mathbf{a}_i))\}_{i=1}^{n} \), and kernel hyperparameters are optimized via marginal likelihood maximization. Based on this training data, the GP posterior at any candidate input \( \mathbf{a} \) is given by:
\[
U(\mathbf{a}) \mid \mathcal{D}_n \sim \mathcal{N}(\mu(\mathbf{a}), \sigma^2(\mathbf{a})),
\]
where \( \mu(\mathbf{a}) \) is the predictive mean and \( \sigma^2(\mathbf{a}) \) reflects the model's uncertainty. This variance term is critical for guiding exploration in the acquisition function.

The optimization process is initialized with \( N_0 \) samples drawn from a uniform grid over the 2D normalized input space. These initial samples provide diverse coverage, enabling the GP to form a stable prior before the sequential search begins.

Given the presented GP model, at each iteration of Bayesian Optimization, we select the next configuration that maximizes the \emph{acquisition function}. In particular, we have: 
\begin{equation}
\mathbf{a}_{n+1} = \arg\max_{\mathbf{a} \in [0,1]^2} \alpha(\mathbf{a}), 
\end{equation}
The selected input \( \mathbf{a}_{n+1} \) is then de-normalized, and the split layer component is rounded to the nearest valid integer. This configuration is evaluated on the actual system, and the resulting observation is added to the dataset \( \mathcal{D}_{n+1} \). The GP surrogate is subsequently updated with this new data point. This iterative process continues until either convergence criteria are met or the evaluation budget is exhausted. Given the iterative nature of the optimization process, a key challenge in leveraging Bayesian Optimization lies in defining an appropriate acquisition function. We next introduce our novel acquisition function tailored for the constrained split inference model. 

\subsection{Hybrid Acquisition Function}
To select the next evaluation point under energy and delay constraints, we define a composite acquisition function that balances three key objectives: maximizing predicted utility, ensuring constraint feasibility, and promoting solution stability. The acquisition function is formulated as:
\begin{equation}
\alpha(\mathbf{a}) = 
\lambda_{\text{ei}}  \alpha_{\text{ei}}(\mathbf{a}) +
\lambda_{\text{ucb}}  \alpha_{\text{ucb}}(\mathbf{a}) -
\lambda_{\text{g}}  \alpha_{\text{grad}}(\mathbf{a}) -
\lambda_{\text{p}} \alpha_{\text{penalty}}(\mathbf{a}),
\label{eq:acquisition}
\end{equation}
where \( \mathbf{a} = [\tilde{P}_t, \tilde{l}] \in [0,1]^2 \) denotes the normalized input vector. Each term in Eq.~\eqref{eq:acquisition} serves a specific purpose. In particular, \( \alpha_{\text{ei}} \) promotes exploration by favoring areas with high expected improvement, while \( \alpha_{\text{ucb}} \) encourages sampling points with high uncertainty and potential.  
Furthermore, \( \alpha_{\text{grad}} \) penalizes regions with steep gradients to enhance stability.  Finally, \( \alpha_{\text{penalty}} \) applies soft penalties for violating energy or delay constraints.
The corresponding weights \( \lambda_{\text{ei}}, \lambda_{\text{ucb}}, \lambda_{\text{g}}, \lambda_{\text{p}} \) control the trade-offs among these competing goals. Next, we provide details for each of these terms.

\noindent\textbf{Expected Improvement.}
The expected improvement (EI) term \( \alpha_{\text{ei}}(\mathbf{a}) \) promotes exploration by favoring regions where the predicted utility is likely to exceed the current best feasible value. It encourages evaluating points that are promising yet underexplored. The EI is defined as:
\begin{equation}
\alpha_{\text{ei}}(\mathbf{a}) = \mathbb{E}\left[\max(0, \mu(\mathbf{a}) - U^*)\right],
\end{equation}
where \( U^* = \max\{U(\mathbf{a}_i) \mid \mathbf{a}_i \in \mathcal{D}_n \cap \mathcal{F} \} \) denotes the best observed utility among all feasible points in the dataset.

\noindent\textbf{Upper Confidence Bound.}
The upper confidence bound (UCB) term \( \alpha_{\text{ucb}}(\mathbf{a}) \) encourages sampling points with both high predicted utility and high model uncertainty. This helps the optimizer explore regions that are uncertain but potentially valuable. The UCB is defined as:
\begin{equation}
\alpha_{\text{ucb}}(\mathbf{a}) = \mu(\mathbf{a}) + \beta \cdot \sigma(\mathbf{a}),
\end{equation}
where \( \mu(\mathbf{a}) \) and \( \sigma(\mathbf{a}) \) are the GP posterior mean and standard deviation, and \( \beta \) controls the exploration-exploitation trade-off.

\noindent\textbf{Stability-Promoting Gradient Penalty.} To enhance the efficiency of the GP, we propose using the gradient penalty term \( \alpha_{\text{grad}}(\mathbf{a}) \) that penalizes regions where the predicted utility changes rapidly with respect to the inputs. Such regions tend to be unstable and sensitive to small perturbations. By discouraging high gradient norms, this term promotes robustness in the selected configuration. This penalty term is defined as:
\begin{equation}
\alpha_{\text{grad}}(\mathbf{a}) = \| \nabla \mu(\mathbf{a}) \|.
\end{equation}

\noindent\textbf{Constraint Penalty.}
The constraint penalty term \( \alpha_{\text{penalty}}(\mathbf{a}) \) softly penalizes configurations that violate energy or delay constraints. This ensures that the acquisition function prioritizes feasible regions while still allowing limited exploration near constraint boundaries. The penalty is defined as:
\begin{align}
\alpha_{\text{penalty}}(\mathbf{a}) =\,
&\left( E_c(\mathbf{a}) + E_t(\mathbf{a}) - E_{\max} \right)^+ \notag \\
+\, &\left( \tau_c^{\text{MD}}(\mathbf{a}) + \tau_t(\mathbf{a}) + \tau_c^{\text{S}}(\mathbf{a}) - \tau_{\max} \right)^+,
\end{align}
where \( (x)^+ = \max(0, x) \) denotes the ReLU operator, used here to apply penalties only when constraints are violated.
\noindent\textbf{Feasible Region.} The feasible region \( \mathcal{F} \) consists of all configurations that satisfy both energy and delay constraints. It is defined as:
\begin{equation}
\mathcal{F} = \left\{ \mathbf{a} \mid E_c(\mathbf{a}) + E_t(\mathbf{a}) \leq E_{\max}, \quad \tau_{\text{total}}(\mathbf{a}) \leq \tau_{\max} \right\},
\end{equation}
where \( \tau_{\text{total}} = \tau_c^{\text{MD}} + \tau_t + \tau_c^{\text{S}} \) is the end-to-end delay, including local computation, transmission, and server-side processing. All terms are computed analytically based on the system model.

%We treat utility evaluations as deterministic but subject to variability due to wireless channel fluctuations. Although the utility function \( U(\mathbf{a}) \) is a black box, the constraint terms \( E_c, E_t, \tau \) are fully known and differentiable. The acquisition combines UCB and EI to balance exploration and exploitation. The gradient term encourages convergence to stable regions, and the penalty discourages infeasible configurations. We schedule each component’s influence across optimization steps, detailed in the next section.

\begin{figure}[h]
    \centering
    \includegraphics[width=0.9\linewidth]{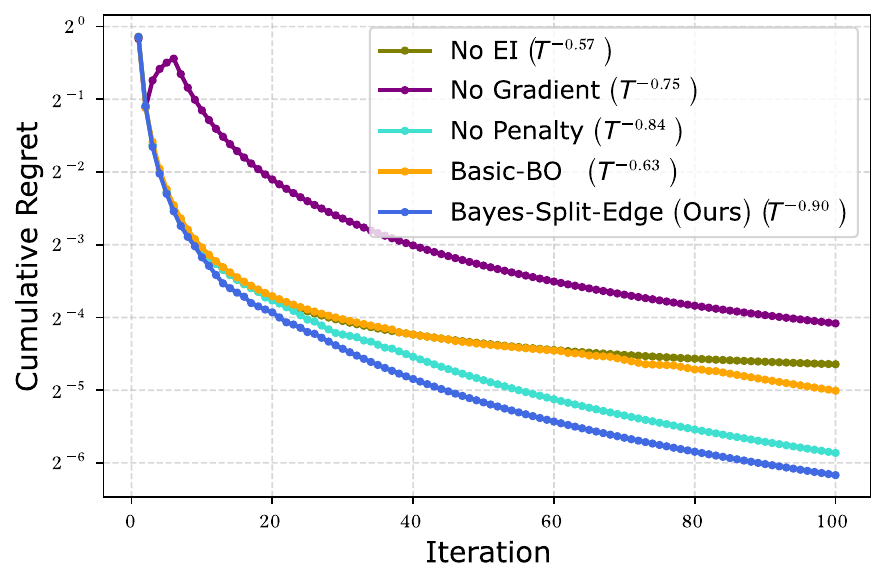}
    \caption{Bayes-Split-Edge components showing cumulative regret over iterations. Our complete method achieves the fastest convergence (\(\mathcal{O}(T^{-0.90})\)). Y-axis scaled in powers of two for readability.}
    \label{fig:ablation}
\end{figure}

\noindent\textbf{Adaptive Weight Scheduling.}
To balance exploration and exploitation over time, we apply exponential decay to selected acquisition weights. Specifically. we define: 
% \begin{align*}
% \lambda_{\text{base}}(t) &= \lambda_{\text{base}}^{(0)} \left( \frac{\lambda_{\text{base}}^{(T)}}{\lambda_{\text{base}}^{(0)}} \right)^t, \\
% \lambda_{\text{g}}(t)     &= \lambda_{\text{g}}^{(0)} \left( \frac{\lambda_{\text{g}}^{(T)}}{\lambda_{\text{g}}^{(0)}} \right)^t,
% \end{align*}
\begin{align*}
\lambda_{\text{base}}(t) &= \lambda_{\text{base}}^{(0)} \left( \frac{\lambda_{\text{base}}^{(T)}}{\lambda_{\text{base}}^{(0)}} \right)^t, 
\\
\lambda_{\text{g}}(t)     &= \lambda_{\text{g}}^{(0)} \left( \frac{\lambda_{\text{g}}^{(T)}}{\lambda_{\text{g}}^{(0)}} \right)^t,
\end{align*}
where \( t = \frac{n}{T-1} \) is the normalized iteration index, \( n \) is the current iteration, and \( T \) is the total number of iterations, and the base weight \( \lambda_{\text{base}} \) controls the influence of utility-driven acquisition terms such as expected improvement (EI) or upper confidence bound (UCB). This allows early-stage exploration to gradually shift toward exploitation as the GP surrogate becomes more accurate. The penalty weight \( \lambda_{\text{p}} \) remains constant to consistently enforce constraint feasibility throughout the optimization process. Algorithm \ref{alg:bo} presents our proposed algorithm. We explore the effect of each component in the section \ref{sec:results} (See Fig.~\ref{fig:ablation}).

% \subsection{Optimization Procedure}

% At each iteration of Bayesian Optimization, we select the next configuration by maximizing the acquisition function:
% \begin{equation}
% \mathbf{a}_{n+1} = \arg\max_{\mathbf{a} \in [0,1]^2} \alpha(\mathbf{a}).
% \end{equation}
% The selected input \( \mathbf{a}_{n+1} \) is then de-normalized, and the split layer component is rounded to the nearest valid integer. This configuration is evaluated on the actual system, and the resulting observation is added to the dataset \( \mathcal{D}_{n+1} \). The GP surrogate is subsequently updated with this new data point. This iterative process continues until either convergence criteria are met or the evaluation budget is exhausted.

\begin{algorithm}[H]
\caption{Split Edge: Bayesian Resource-Optimizer}\label{alg:bo}
\begin{algorithmic}[1]
\Require Initial dataset \( \mathcal{D}_0 = \{(\mathbf{a}_i, U(\mathbf{a}_i))\}_{i=1}^{N_0} \)
\Require GP prior \( \mathcal{GP}(\mu_0, k_0) \)
\Require Evaluation budget \( T \), early stop threshold \( N_{\max} \)
\Require Acquisition weights: \( \lambda_{\text{base}}^{(0)}, \lambda_{\text{base}}^{(T)}, \lambda_{\text{g}}^{(0)}, \lambda_{\text{g}}^{(T)}, \lambda_{\text{p}} \)

\State \textbf{Initialize:}
\State Fit GP on \( \mathcal{D}_0 \)
\State Set \( \mathbf{a}^* \leftarrow \arg\max_{\mathbf{a}_i \in \mathcal{D}_0 \cap \mathcal{F}} U(\mathbf{a}_i) \)
\State Set counter \( n_c \leftarrow 0 \)

\For{iteration \( n = N_0+1 \) to \( T \)}
    \State Normalize input domain \( \mathbf{a} \in [0,1]^2 \)
    \State Compute normalized index \( t \leftarrow \frac{n - N_0}{T-1} \)
    \State Update weights: \(\lambda_{\text{base}}(t)\) and \(\lambda_{\text{g}}(t) \)
    \State Compute GP posterior: \( \mu(\mathbf{a}), \sigma(\mathbf{a}), \nabla \mu(\mathbf{a}) \)
    
    \State Evaluate acquisition function:
    \begin{align*}
    \alpha(\mathbf{a}) =\,
    &\lambda_{\text{base}} \cdot \alpha_{\text{ei}}(\mathbf{a}) +
     \lambda_{\text{base}} \cdot \alpha_{\text{ucb}}(\mathbf{a}) \\
    &- \lambda_{\text{g}} \cdot \|\nabla \mu(\mathbf{a})\| -
     \lambda_{\text{p}} \cdot \alpha_{\text{penalty}}(\mathbf{a})
    \end{align*}
    
    \State Select next configuration: \( \mathbf{a}_n \leftarrow \arg\max_{\mathbf{a} \in \mathcal{A}} \alpha(\mathbf{a}) \)
    \State Evaluate utility: \( U(\mathbf{a}_n) \)
    \State Update dataset: \( \mathcal{D}_n \leftarrow \mathcal{D}_{n-1} \cup \{(\mathbf{a}_n, U(\mathbf{a}_n))\} \)
    
    \If{\( \mathbf{a}_n = \mathbf{a}^* \)}
        \State \( n_c \leftarrow n_c + 1 \)
        \If{\( n_c \geq N_{\max} \)} 
            \State \textbf{break} 
        \EndIf
    \Else
        \State \( \mathbf{a}^* \leftarrow \mathbf{a}_n \); \( n_c \leftarrow 0 \)
    \EndIf
    
    \State Refit GP on \( \mathcal{D}_n \)
\EndFor

\State \textbf{return} Best configuration \( \mathbf{a}^* = [P_t^*, l^*] \), utility \( U(\mathbf{a}^*) \)
\end{algorithmic}
\end{algorithm}

\subsection{Regret Analysis}
We consider the standard cumulative regret definition:
\begin{equation}
R_T := \sum_{t=1}^T \left[ U(x^\star) - U(x_t) \right],
\end{equation}
where \( x^* \) is the global optimum in the feasible region \( \mathcal{X}_\delta \).

Our hybrid acquisition function combines UCB, EI, gradient stability, and feasibility penalty terms, with adaptive weights. Following the Gaussian Process bandit framework in~\cite{srinivas2010gaussian, chowdhury2017kernelized}, and extending recent analyses on constrained Bayesian optimization~\cite{bogunovic2016truncated, gardner2014bayesian}, we obtain the following bound:

\begin{theorem}[Cumulative Regret]
Assume the objective lies in the RKHS of a Matérn kernel, constraints are Lipschitz continuous, and the optimum is well-separated from the boundary. Then the cumulative regret of our method satisfies:
\begin{equation}
R_T = \mathcal{O} \left( \sqrt{T \cdot \gamma_T^{(\delta)} } \right),
\end{equation}
where \( \gamma_T^{(\delta)} \) is the information gain over the feasible region.
\end{theorem}

For small feasible sets (i.e., \( |\mathcal{X}_\delta|/|\mathcal{X}| \leq T^{-1/2} \)), this reduces to:
\begin{equation}
R_T = \mathcal{O} \left( \sqrt{T \cdot \log^{d+1} T} \right).
\end{equation}
\begin{proof}
    The proof follows from confidence bounds and information-theoretic arguments in~\cite{srinivas2010gaussian}, extended with the feasibility margin conditions from~\cite{bogunovic2016truncated} and the hybrid acquisition decomposition from our framework. Details are omitted here for brevity.
\end{proof}

\section{Performance Evaluation}\label{sec:results}
In this section, we present numerical and experimental results to compare the performance of our proposed solution against several baseline algorithms.

% \begin{figure}[t]
%     \centering
%     \includegraphics[width=0.95\linewidth]{figures/regret_comparison.pdf}
%     \caption{Cumulative regret comparison for split-inference optimization. Bayes-Split-Edge achieves a near-power-law convergence rate of \(\mathcal{O}(T^{-0.85})\), substantially faster than the \(\mathcal{O}(T^{-0.43})\) of Basic-BO, underscoring the advantages of our hybrid acquisition function in Eq.~\eqref{eq:acquisition}. \emph{ For improved readability, the y-axis is scaled in powers of two without logarithmic transformation; regret values are reported in their native units.}}

%     \label{fig:regret-comparison}
% \end{figure} 

\subsection{Simulation Setup}
We simulate per-sample inference in a realistic edge–server split execution setting. The {edge device} is a Raspberry Pi 4 (4 cores, 1.8\,GHz), and the {edge server} is a Mac M4 (10 cores, 4.5\,GHz). 
\begin{figure}[H]
\centering
\includegraphics[width=0.22\textwidth]{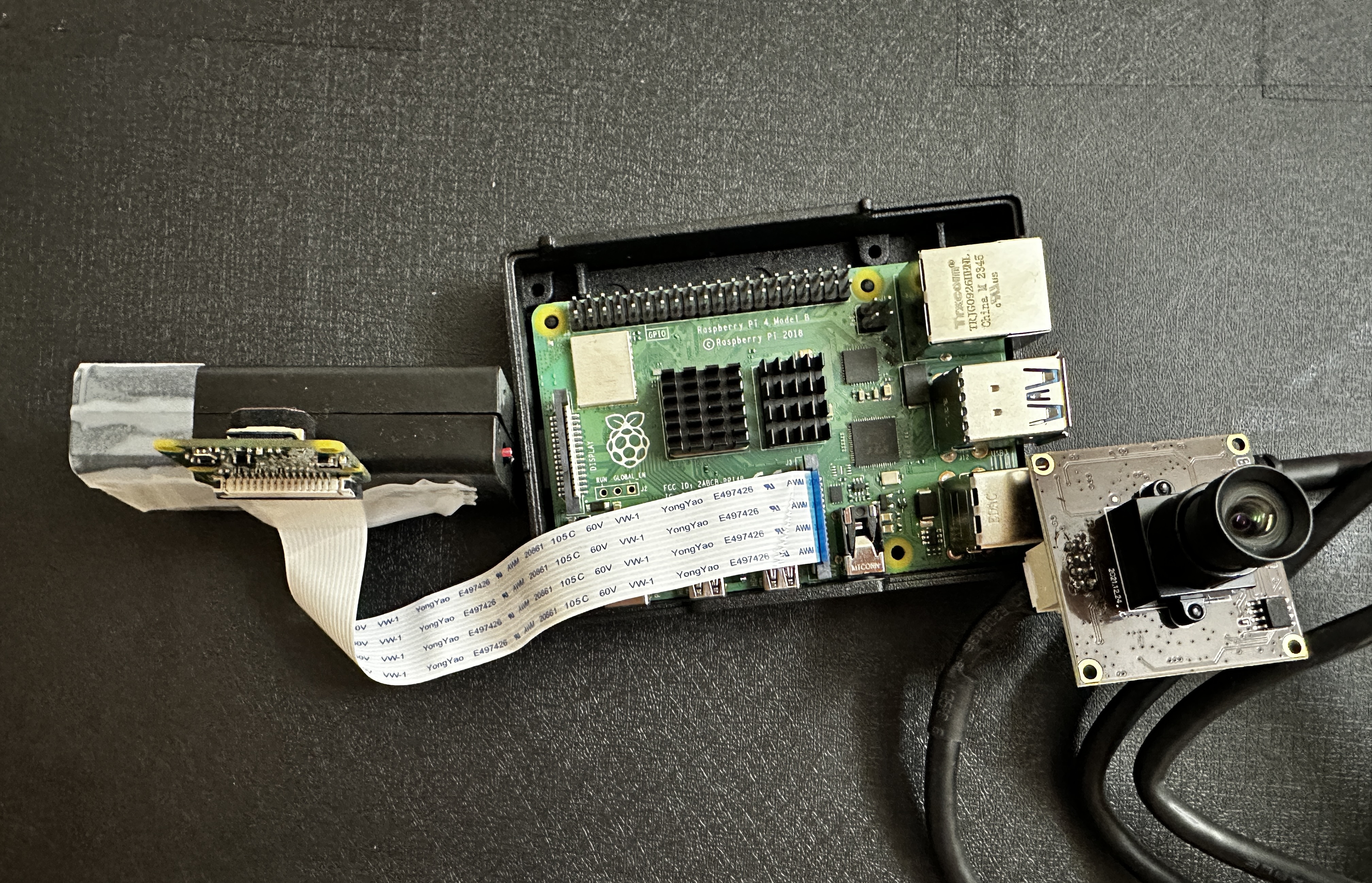}
\includegraphics[width=0.24\textwidth]{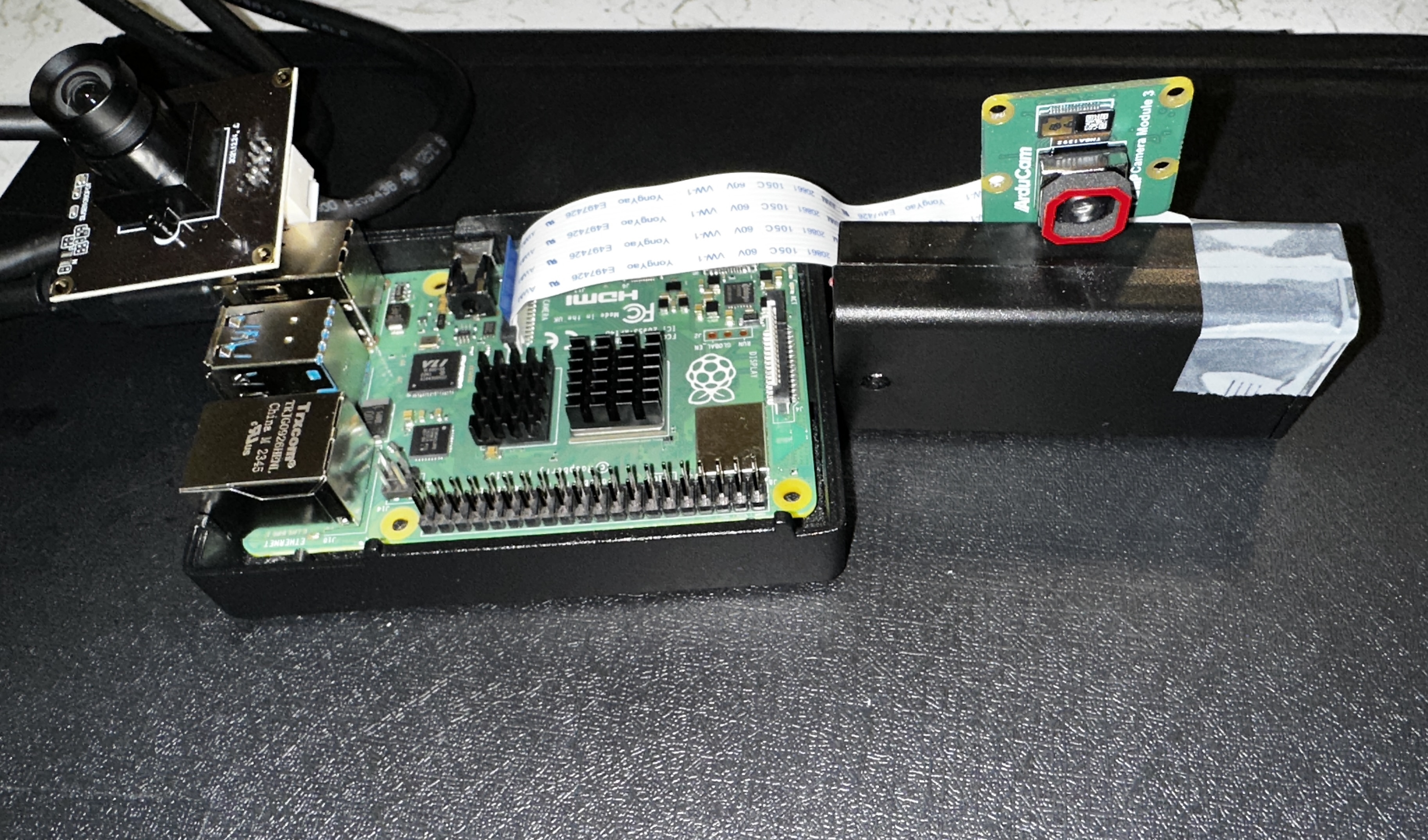}
\caption{Raspberry Pi 4 experimental setup demonstrating real-world edge constraints. The limited computational resources (4GB RAM, ARM Cortex-A72) and thermal constraints (visible heat sinks) directly motivate our constraint-aware optimization approach. Camera module and wireless connectivity represent typical split-inference deployment scenarios where energy and latency budgets are critical.}
\label{fig:hardware_setup}
\end{figure}
Server-side energy is assumed unconstrained and thus not modeled. We evaluate performance over multiple wireless channel traces from the \textit{mMobile} dataset \cite{mmobile}, specifically using the Outdoor, with link length of 30 \(m\), resolution	of 0.6\(m\), tracing 45 points with blockage. %These traces provide real-world mobility and fading effects, and we use them to verify performance robustness—not to average out stochasticity.\blue{how we used the channel info: fast fading} Each algorithm is tested under varying conditions to confirm its adaptability. 
These traces capture real-world mobility and fast fading effects, which we employ to assess performance robustness rather than to average over stochasticity. The reported results derive from a single channel realization; we evaluate each algorithm to determine the one that achieves convergence most rapidly or attains the optimal solution most expeditiously within that realization.

The wireless link uses a bandwidth of $B = 240{,}000 \times 256 \times 0.8$\,Hz, to model practical subcarrier allocation in an OFDM system. The noise power spectral density is set to $-147$\,dBm/Hz. 

The computation task is image classification using \textbf{VGG19} \cite{vgg} evaluated on the \textbf{ImageNet-Mini} dataset \cite{kaggle_imagenet_mini} consisted of $1,000$ samples across $100$ classes. Split layers are selectable from layer 1 through 37. We use batch size 1 to model real-time inference. $\kappa = 10^{-29}$ models the device's energy coefficient, and $f = 1.8$\,GHz is the device’s CPU frequency. FLOPs per layer are obtained from the model architecture.
Inference is performed in FP32 precision to maintain numerical stability and accuracy. For cases where latency becomes infeasible under strict resource limits, we apply a deadline-based truncation approach. This method resembles dropout by stopping the input data stream once the deadline is reached, which skips the remaining tail layers and avoids exceeding resource bounds.
%Inference is done in FP32 precision \blue{add more details about the compression}. To handle infeasible latency cases, we simulate a dropout-like method which skips tail layers that would exceed resource constraints.

All algorithms are evaluated under strict budgets: maximum energy of 5\,J and latency of 5\,s per inference. We report accuracy, total energy, total delay, and number of function evaluations as evaluation metrics.
\begin{table*}[h]
\centering
\caption{Performance comparison of optimization methods on split-inference task.Bayes-Split-Edge matches the global optimum found by exhaustive search using only 20 iterations (1800$\times$ fewer evaluations), while other methods either require significantly more samples or converge to suboptimal solutions.}
\begin{tabular}{lcccccc}
\toprule
\textbf{Algorithm} & \textbf{Max Iterations} & \textbf{Split Layer} & \textbf{Transmit Power (Watt)} & \textbf{Accuracy (\%)} & \textbf{Energy (J)} & \textbf{Delay (s)} \\
\midrule
\textbf{Bayes-Split-Edge (Ours)}   & \textbf{20}    & \textbf{7} & \textbf{0.38} & \textbf{87.50} & \textbf{1.53} & \textbf{5.00} \\
Basic-BO                   & 48             & 7          & 0.4          & 85.94          & 1.53          & 5.00 \\
Exhaustive Search            & 36036           & 7          & 0.35--0.39    & 87.50          & 1.53          & 5.00 \\
Direct Search                & 80             & 7          & 0.38          & 87.50          & 1.53          & 5.00 \\
CMA-ES                       & 32             & 2          & 0.10          & 84.38          & 0.11          & 3.75 \\
Random Search                & 300            & 3          & 0.28          & 84.38          & 0.61          & 4.01 \\
RL (PPO)       & 100            & 5          & 0.17          & 84.38          & 1.02          & 4.39 \\
Transmit-First               & 1              & 1          & 0.50          & 84.38          & 0.14          & 3.31 \\
Compute-First                & 1              & 7          & 0.34          & 84.38          & 1.53          & 5.00 \\
\bottomrule
\end{tabular}
\label{tab:split_comparison}
\end{table*}

\begin{figure*}[t]
    \centering
    \includegraphics[width=0.98\linewidth]{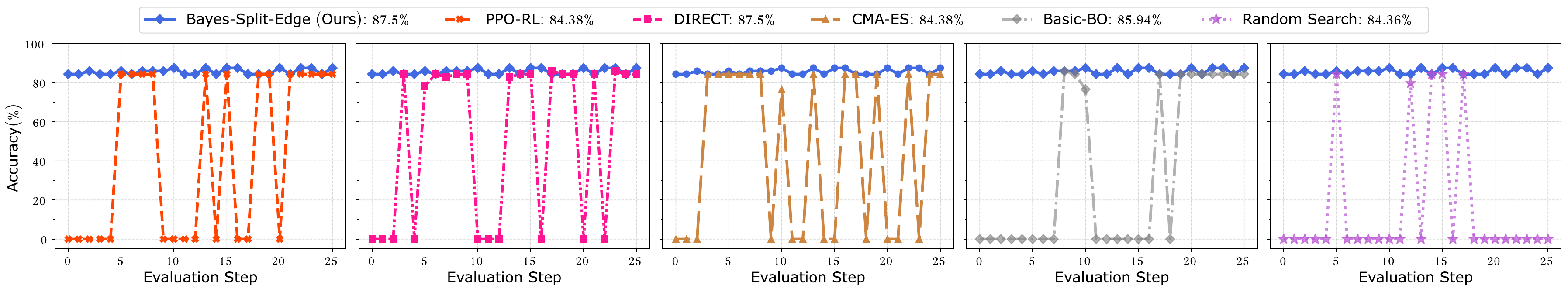}
    \caption{
    \textbf{Accuracy vs. evaluation step} for six split-inference strategies under a 5(J)-5(s) resource budget.  Bayes-Split-Edge (blue) consistently operates in the feasible regime, never collapsing to zero, and maintains accuracy above 80\%.  It converges to its peak (87.5 \%) in fewer than 20 function evaluations, outpacing PPO-RL (red), DIRECT (magenta), CMA-ES (brown), Basic-BO (gray), and Random Search (orchid), all of which either require more evaluations to reach their best accuracy or repeatedly violate feasibility.
    }
    \label{fig:bo_vs_rl_accuracy}
\end{figure*}
\subsection{Baseline Algorithms}\label{sec:baselines}
To evaluate Split-Edge, we compare it against eight baseline algorithms: Vanilla Bayesian Optimization, Exhaustive Search, Direct Search, CMA-ES, Random Search, Reinforcement Learning, and Transmit-First and Compute heuristic algorithms. 

Exhaustive Search method performs a complete search over the joint space of transmission power and split layer configurations. Assuming the split layer \( l \in \{1, 2, \dots, L\} \) and the transmission power \( P_t \in \mathcal{P} \) is quantized into \( |\mathcal{P}| \) discrete levels, Exhaustive Search evaluates all \( L \times |\mathcal{P}| \) configurations. For each pair \( (l, P_t) \), it computes the corresponding accuracy, delay, and energy, then selects the best feasible configuration based on utility. While this approach guarantees global optimality, the total number of function evaluations grows linearly with both the number of split points and the granularity of the power levels, which results in the 
\text{total evaluations} of \( \mathcal{O}(L \cdot |\mathcal{P}|).\)
Due to its high computational cost, it is only used as an offline benchmark and is not viable for deployment in real-time or adaptive scenarios.

The Standard-BO optimizes the black-box utility function using a standard acquisition function, such as Upper Confidence Bound (UCB) or Expected Improvement (EI), over the input space of split layer and transmission power. These functions are agnostic to feasibility constraints like energy or delay, focusing solely on maximizing expected utility or exploration potential. As a result, the Standard-BO frequently selects infeasible configurations, particularly in the early stages, leading to inefficient sampling. The number of function evaluations is proportional to the total budgeted rounds \( T \), with theoretical convergence characterized by sublinear regret bounds \( \mathcal{O}(\sqrt{T}) \). However, in constrained and structure-rich environments, this approach often fails to exploit problem-specific priors, resulting in slow or suboptimal convergence.

We include the classical DIRECT algorithm \cite{jones1993lipschitzian} as a gradient-free baseline. DIRECT begins by evaluating the objective, negative accuracy, at the center of the search domain, with configurations exceeding the 5\,J energy or 5\,s latency budgets assigned zero accuracy. At each iteration, it selects potentially optimal rectangles based on Lipschitz constant estimates, divides their longest dimension into three equal parts, and evaluates the centers of the new subregions. We cap the search at 100 evaluations and terminate early if accuracy does not improve for 20 consecutive trials. Despite using neither gradients nor surrogate models, DIRECT reliably identifies high-utility, feasible configurations in our mixed-integer search space.

We include CMA-ES \cite{hansen2001completely} as an adaptive, gradient-free baseline. CMA-ES maintains a multivariate normal distribution over normalized transmit power and split-layer index and samples a population of $10$ candidates each generation. Sampled layer values are denormalized and rounded to the nearest integer; any configuration that violates the energy or delay constraint is scored with zero accuracy. After evaluating each generation, CMA-ES updates its mean and covariance matrix using its standard self-adaptation rules. We cap the search at 300 evaluations and terminate early if accuracy does not improve for 20 consecutive samples. Although CMA-ES is effective at guiding the search toward high-accuracy configurations without gradients, it frequently evaluates options that violate our energy or latency limits. This inefficiency underscores the value of constraint-aware methods such as Split-Edge.

We include Random Search as a simple, gradient-free baseline that uniformly samples 300 configurations across the split layer and transmit power bounds. Each sampled pair $(P_t, l)$ is denormalized and rounded before evaluation; any configuration that exceeds the 5\,J energy or 5\,s latency limits is assigned zero accuracy. Because Random Search ignores both past evaluations and problem structure, it occasionally discovers high-accuracy, feasible solutions but generally exhibits poor sample efficiency and frequent constraint violations. This behavior underscores the benefit of more informed, constraint-aware methods such as Split-Edge.

We adopt PPO as a reinforcement learning baseline, motivated by Zhang et al.\cite{zhang2024proximal}, who successfully applied PPO to joint offloading and power allocation in multi-access edge computing. We formulate split inference as an MDP where the agent observes the previous normalized transmit power and split-layer index as state. At each step, the agent outputs a continuous action in $[0,1]^2$ representing new power and layer selections, which we denormalize to physical values and round layer indices to integers. The environment returns a reward equal to inference accuracy, with a $-5$ penalty for configurations violating the 5\,J energy or 5\,s latency constraints. State transitions add Gaussian noise ($\sigma = 0.01$) to the current action, providing a simple exploration mechanism. We train the policy for 100 timesteps using standard PPO hyperparameters (entropy coefficient 0.05, learning rate $3 \times 10^{-4}$) and evaluate over 100 deterministic rollouts. Despite utilizing all 100 function evaluations, the severely constrained training budget and noisy dynamics prevent meaningful policy learning, with PPO consistently underperforming Split-Edge.

We also implement two greedy heuristics that prioritize single resources. Transmit-First sets \(P_t = P_{max}\) and searches for the deepest feasible split layer, decrementing power if no valid configuration exists.
Compute-First fixes the deepest split layer and finds maximum feasible transmit power, backing off layers incrementally if infeasible. Both require linear search and illustrate the suboptimality of single-resource optimization under joint constraints.

\begin{figure*}[h]
    \centering
    \includegraphics[width=0.98\linewidth]{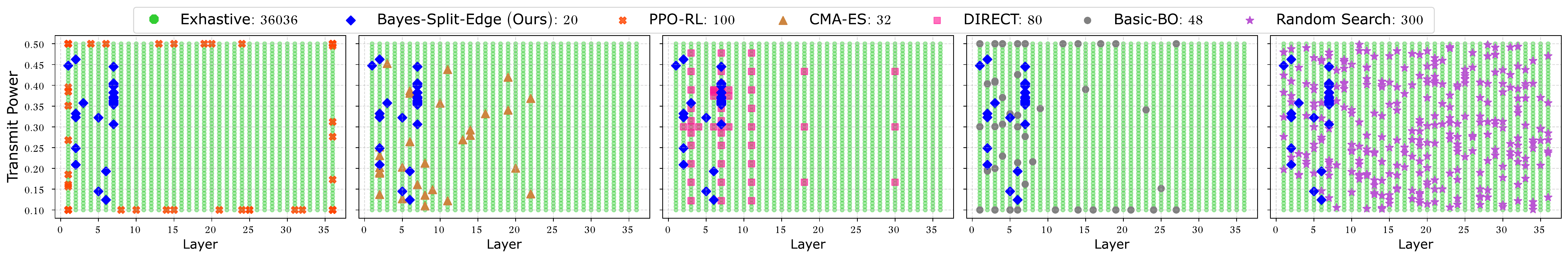}
    \caption{
     \textbf{Split-layer and transmit-power search space} under a 5(J)-5(s) budget. Exhaustive search (36,036 evaluations) provides ground truth. Our Bayes-Split-Edge method (blue, 20 evaluations) efficiently finds the global optimum by only searching within feasibility regions, while baselines, PPO-RL (100), CMA-ES (32), DIRECT (80), Basic-BO (48), and Random Search (300), waste samples in infeasible regions and converge poorly.}    \label{fig:search_space}
\end{figure*}

\subsection{Evaluation Results}
\begin{figure}[h]
    \centering

    % First subplot (a)
    \begin{minipage}{0.95\linewidth}
        \centering
        \includegraphics[width=\linewidth]{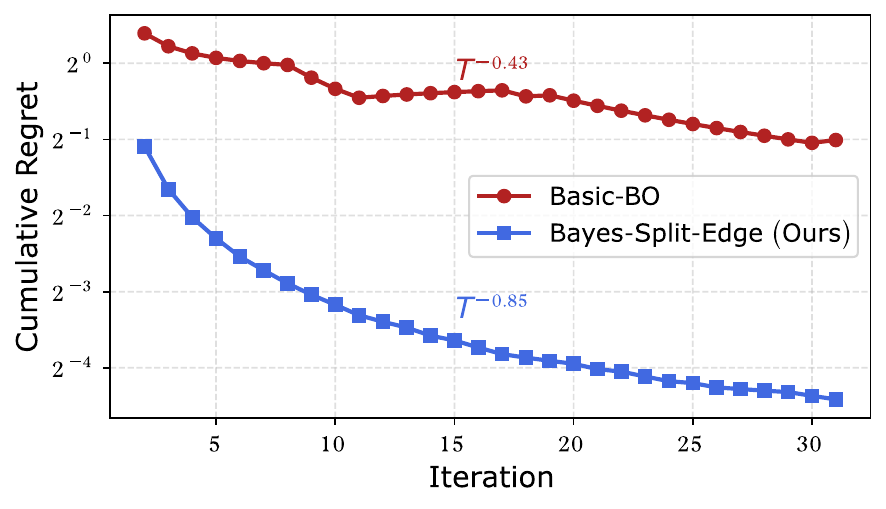}
        \vspace{0.3em}
        \textbf{(a)} \small ImageNet-Mini/VGG19
        \label{fig:regret-comparison-row1}
    \end{minipage}

    % Second subplot (b)
    \begin{minipage}{0.95\linewidth}
        \centering
        \includegraphics[width=\linewidth]{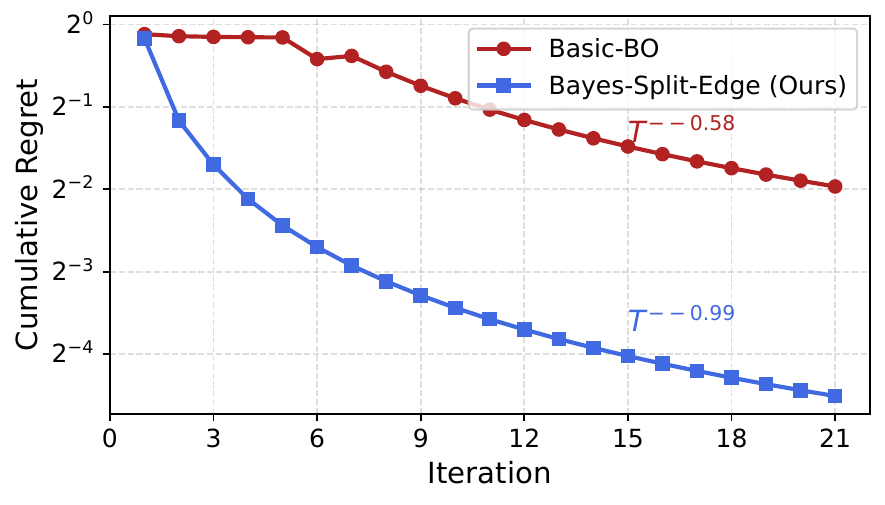}
        \vspace{0.3em}
        \textbf{(b)} \small ImageNet-Tiny/ResNet-101
        \label{fig:regret-comparison-row2}
    \end{minipage}

    \vspace{0.8em}

    \caption{Cumulative regret comparison for split-inference optimization across two dataset/model pairs. 
    Bayes-Split-Edge consistently achieves faster and more efficient convergence than alternative methods, 
    demonstrating the efficacy and model-agnostic nature of our hybrid acquisition approach. 
    The method can be adopted for any neural network architecture without modification.\emph{ For improved readability, the y-axis is scaled in powers of two without logarithmic transformation; regret values are reported in their native units.} 
    }
    \label{fig:regret-comparison-panel}
\end{figure}

% Figure~\ref{fig:regret-comparison} presents the normalized regret $
% \bar{R}_T := \frac{1}{T} \sum_{t=1}^{T} \left( U(x^\star) - U(x_t) \right).$ In this plot, the y-axis is log-scaled to emphasize the rate of regret decay. From the results, we observe that 
% while both methods exhibit sublinear growth, consistent with theoretical bounds of the form \( R_T = \mathcal{O}(\sqrt{T \gamma_T}) \), Split-Edge demonstrates a steeper decline.
% This indicates faster convergence for Split-Edge, driven by hybrid acquisition and feasible-region exploitation.

Figure~\ref{fig:regret-comparison-panel} provides theoretical validation of our method's superior convergence properties through cumulative regret analysis. The plot presents the normalized regret $\bar{R}_T := \frac{1}{T} \sum_{t=1}^{T} \left( U(x^\star) - U(x_t) \right)$ with a log-scaled y-axis to emphasize the rate of regret decay. While both methods exhibit sublinear growth, consistent with theoretical bounds of the form $R_T = \mathcal{O}(\sqrt{T \gamma_T})$, Bayes-Split-Edge demonstrates a dramatically steeper decline. This indicates faster convergence for Split-Edge, driven by hybrid acquisition and feasible-region exploitation.

Our method achieves near-linear regret decay at $\mathcal{O}(T^{-0.85})$, approaching the theoretical optimum of $\mathcal{O}(T^{-1})$ for constrained optimization problems. This represents a substantial improvement over Basic-BO's $\mathcal{O}(T^{-0.43})$ convergence rate—nearly doubling the convergence exponent. The superior regret bounds stem from two key algorithmic contributions in our hybrid acquisition function (Eq.~\ref{eq:acquisition}): constraint-aware sampling eliminates wasted evaluations on infeasible configurations, and gradient estimation provides directional guidance that accelerates convergence toward the global optimum.

The faster convergence for Bayes-Split-Edge is driven by hybrid acquisition and feasible-region exploitation, which fundamentally changes the optimization dynamics. While Basic-BO requires $\mathcal{O}(\epsilon^{-2.33})$ iterations to achieve regret threshold $\epsilon$, Bayes-Split-Edge needs only $\mathcal{O}(\epsilon^{-1.18})$  iterations. This theoretical advantage translates directly to the empirical results in Table~\ref{tab:split_comparison} and ~\ref{fig:bo_vs_rl_accuracy}, where we observe 2.4$\times$ faster convergence to optimal accuracy for Bayes-Split-Edge method, whereas the Basic-BO even with more iterations does not converge to the optimal accuracy.
% \textbf{Normalized Cumulative Regret.}
% The normalized regret is:
% \begin{equation}
% \bar{R}_T := \frac{1}{T} \sum_{t=1}^{T} \left( f(x^\star) - f(x_t) \right).
% \end{equation}

Figure~\ref{fig:bo_vs_rl_accuracy} reveals the critical importance of constraint-aware optimization in split-inference planning. Our Bayes-Split-Edge method (blue) maintains consistent 87.5\% accuracy throughout the optimization process, demonstrating perfect constraint satisfaction across all 20 evaluations. This stable performance stems from our constraint-aware acquisition function that explicitly models feasibility, ensuring every sample contributes meaningful information toward the optimization objective.

The baseline methods exhibit fundamentally different behavior patterns that highlight their limitations. PPO-RL (orange dashed) shows the most dramatic instability, with accuracy oscillating between 0\% and 85\% due to frequent constraint violations. These catastrophic drops occur when the policy explores infeasible action spaces, effectively wasting 30--40\% of evaluation budget on unusable configurations. CMA-ES (brown) displays similar oscillatory behavior but with slightly better recovery patterns, though still suffering from 15--20 constraint violations across 32 evaluations. DIRECT (magenta) achieves the same final accuracy as our method (87.5\%) but requires 4$\times$ more evaluations (80 vs 20) due to its constraint-agnostic rectangular partitioning strategy that systematically explores infeasible regions.

Basic-BO (gray) presents the most interesting comparison, maintaining relative stability while achieving 85.94\% accuracy which is slightly lower than our method despite using 2.4$\times$ more evaluations (48 vs 20). The key difference lies in exploration efficiency: Basic-BO's gradient-free acquisition leads to conservative sampling that avoids catastrophic failures but also misses the true optimum. Our gradient-enhanced acquisition strikes the optimal balance, directing search toward high-reward regions while respecting constraints.

These results demonstrate that constraint violations are not merely inefficient as they fundamentally destabilize the optimization process. The consistent performance of Bayes-Split-Edge across evaluation steps shows that joint constraint-gradient modeling eliminates the exploration-exploitation dilemma in constrained optimization, enabling reliable deployment in resource-critical environments where constraint violations can cause system failures.

\begin{figure}[H]
\centering
\includegraphics[width=0.27\textwidth]{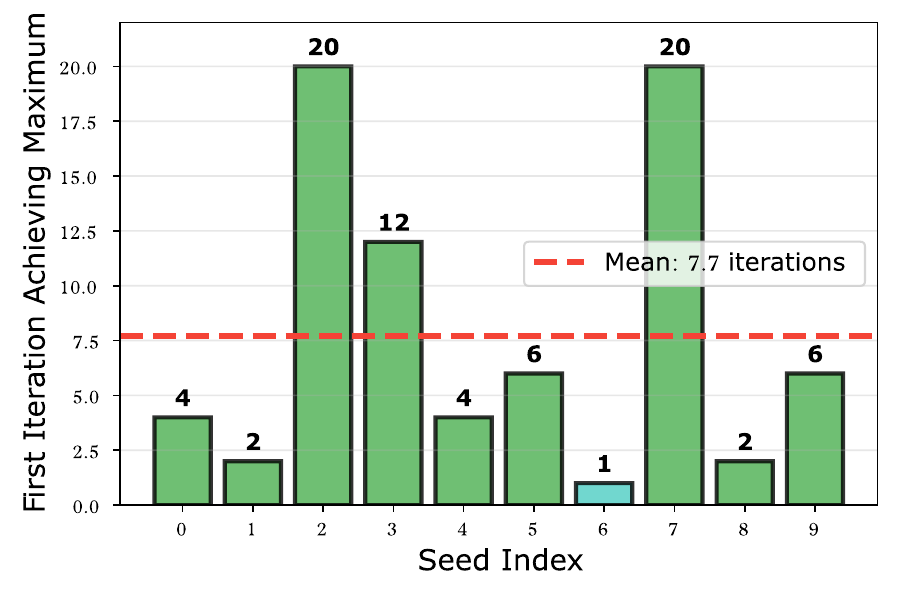}
\includegraphics[width=0.18\textwidth]{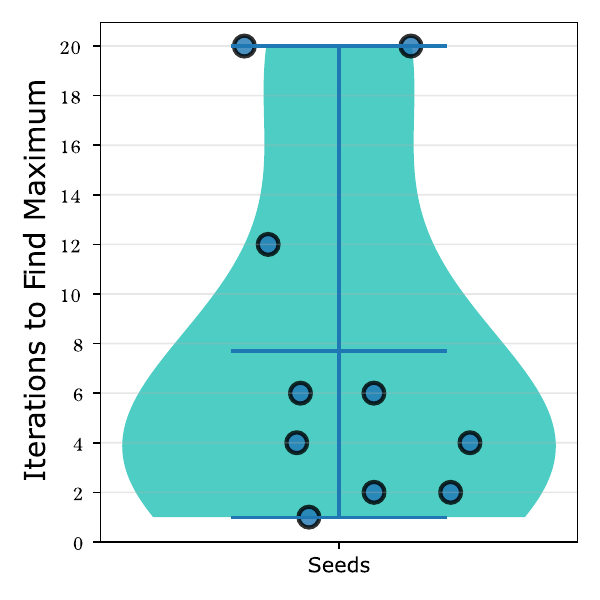}

\caption{Convergence iteration across 10 random seeds for Bayes-Split-Edge. Despite varying convergence speeds, all seeds successfully reach the global optimum accuracy of 87.5\%, below 20 iterations and on average less than 8 iterations.}
\label{fig:random_seed}
\end{figure}

Figure~\ref{fig:search_space} and Table ~\ref{tab:split_comparison} demonstrate the superior sample efficiency and constraint-awareness of Bayes-Split-Edge across the joint split-layer and transmit-power optimization space with the given energy and delay budgets. The green points indicate explored points by the Exhaustive search method. Our method (blue diamonds, $20$ evaluations) exhibits three critical advantages: (1) \textbf{Complete constraint satisfaction}: every sample lies within feasible regions, demonstrating effective constraint-aware acquisition; (2) \textbf{Rapid convergence}: samples concentrate around the global optimum (layers 7, power $\sim$0.35--0.45 W) within the first 10 evaluations one average (see Figure.~\ref{fig:random_seed}); (3) \textbf{Gradient-guided exploration}: the clustered sampling pattern indicates our gradient estimates successfully direct search toward high-reward regions.

We note that Bayes-Split-Edge ensures constraint-aware optimization of both split-layer and transmit-power selection, tailored to the underlying device and network conditions, thereby delivering feasible high-utility solutions across diverse scenarios. In contrast, Compute-First disregards transmit power and deadlines; when activations cannot be fully transmitted, the resulting truncation degrades accuracy. Our approach jointly optimizes the split point and power allocation to mitigate such issues, preserving performance under resource constraints.
In contrast, baseline methods reveal fundamental limitations across multiple dimensions. PPO-RL and Random Search extensively sample infeasible regions, wasting 40--60\% of their evaluation budget on constraint violations leading to 0\% accuracy as in Figure~\ref{fig:bo_vs_rl_accuracy}. CMA-ES shows slow convergence despite constraint-free sampling, requiring the full 32 evaluations to approach the optimum. DIRECT's rectangular partitioning strategy ignores the irregular constraint geometry, which leads to exploration of infeasible space. Basic-BO lacks directional guidance since it only relies on standard acquisition functions, and therefore requires 2.4$\times$ more evaluations to achieve comparable performance.

The results highlight two key insights: first, the 1800$\times$ reduction in samples compared to exhaustive search ($20$ vs. $36,036$) while maintaining optimality demonstrates the dramatic efficiency gains possible through our hybrid acquisition function \ref{eq:acquisition}. Second, the clear separation between our method's concentrated sampling pattern and the scattered baseline approaches shows that constraint-aware acquisition fundamentally changes the optimization dynamics, and transforms intractable search spaces into efficiently navigable landscapes suitable for real-time inference planning.

\section{Conclusion}\label{sec:conclusion}
We presented Bayes-Split-Edge, a constraint-aware Bayesian optimization framework for joint split layer and transmit power selection in collaborative edge inference under fast fading wireless conditions. Our approach combines analytical models of per-layer computation and transmission costs with a hybrid acquisition function that navigates the challenging non-stationary optimization landscape created by rapid channel variations. The proposed hybrid acquisition function balances exploration and exploitation to efficiently find good solutions despite the irregular optimization space caused by different layer architectures. 

We established theoretical regret bounds that guarantee convergence under fast fading conditions, and also validated our approach using a Raspberry Pi 4 edge device, a Mac M4 server, and VGG19 inference on the ImageNet-Mini dataset under tight constraints (5J energy, 5s latency). Our results demonstrate that the proposed method achieves fast convergence (within 20 function evaluations), while maintaining optimal performance. The enabled sample efficiency matters because each evaluation requires running actual inference under real wireless conditions, which costs time and energy on resource-limited devices.  We also showed that {\em Bayes-Split-Edge} consistently outperforms several baselines, including Basic-BO, CMA-ES, DIRECT, PPO-RL, and greedy methods, in classification accuracy. 

Overall, our results indicate that appropriately designed Bayesian optimization can address the stringent temporal constraints of wireless fast fading channels and resource-limited edge devices. This enables a practical real-time collaborative inference in edge computing systems. For future work, we plan to pursue several extensions of this study: (1) handling multiple devices competing for server resources, (2) accounting for server computational limits and concurrent requests, (3) using optimization history and channel patterns to reduce evaluations further, and (4) practical deployment on XR headsets in related applications.

\begin{acks}
The material is based upon work supported in part by NSF grants 1955561,
2323189, 2434113, 2514415, 2032033, 2106150, and 2346528. Any opinions, findings, conclusions, or
recommendations expressed in this material are those of the author(s) and
do not necessarily reflect the views of NSF.
\end{acks}

\bibliographystyle{ACM-Reference-Format}
\bibliography{sample-base}


\begin{thebibliography}{51}

%%% the Web address.

\ifx \showCODEN    \undefined \def \showCODEN     #1{\unskip}     \fi
\ifx \showISBNx    \undefined \def \showISBNx     #1{\unskip}     \fi
\ifx \showISBNxiii \undefined \def \showISBNxiii  #1{\unskip}     \fi
\ifx \showISSN     \undefined \def \showISSN      #1{\unskip}     \fi
\ifx \showLCCN     \undefined \def \showLCCN      #1{\unskip}     \fi
\ifx \shownote     \undefined \def \shownote      #1{#1}          \fi
\ifx \showarticletitle \undefined \def \showarticletitle #1{#1}   \fi
\ifx \showURL      \undefined \def \showURL       {\relax}        \fi
% invisible to TeX
\providecommand\bibfield[2]{#2}
\providecommand\bibinfo[2]{#2}
\providecommand\natexlab[1]{#1}
\providecommand\showeprint[2][]{arXiv:#2}

\bibitem[Abrol and Jha(2016)]%
        {7448820}
\bibfield{author}{\bibinfo{person}{Akshita Abrol} {and} \bibinfo{person}{Rakesh~Kumar Jha}.} \bibinfo{year}{2016}\natexlab{}.
\newblock \showarticletitle{Power Optimization in 5G Networks: A Step Towards GrEEn Communication}.
\newblock \bibinfo{journal}{\emph{IEEE Access}}  \bibinfo{volume}{4} (\bibinfo{year}{2016}), \bibinfo{pages}{1355--1374}.
\newblock
\href{https://doi.org/10.1109/ACCESS.2016.2549641}{doi:\nolinkurl{10.1109/ACCESS.2016.2549641}}


\bibitem[Avasalcai et~al\mbox{.}(2019)]%
        {8812207}
\bibfield{author}{\bibinfo{person}{Cosmin Avasalcai}, \bibinfo{person}{Christos Tsigkanos}, {and} \bibinfo{person}{Schahram Dustdar}.} \bibinfo{year}{2019}\natexlab{}.
\newblock \showarticletitle{Decentralized Resource Auctioning for Latency-Sensitive Edge Computing}. In \bibinfo{booktitle}{\emph{2019 IEEE International Conference on Edge Computing (EDGE)}}. \bibinfo{pages}{72--76}.
\newblock
\href{https://doi.org/10.1109/EDGE.2019.00027}{doi:\nolinkurl{10.1109/EDGE.2019.00027}}


\bibitem[Badnava et~al\mbox{.}(2025)]%
        {BadnavaCH:24a}
\bibfield{author}{\bibinfo{person}{Babak Badnava}, \bibinfo{person}{Jacob Chakareski}, {and} \bibinfo{person}{Morteza Hashemi}.} \bibinfo{year}{2025}\natexlab{}.
\newblock \showarticletitle{Neural-Enhanced Rate Adaptation and Computation Distribution for Emerging {mmWave} Multi-User {3D} Video Streaming Systems}.
\newblock \bibinfo{journal}{\emph{IEEE Trans. Multimedia}} (\bibinfo{date}{Jan.} \bibinfo{year}{2025}).
\newblock
\newblock
\shownote{accepted}.


\bibitem[Bogunovic et~al\mbox{.}(2016)]%
        {bogunovic2016truncated}
\bibfield{author}{\bibinfo{person}{Ilija Bogunovic}, \bibinfo{person}{Jonathan Scarlett}, {and} \bibinfo{person}{Volkan Cevher}.} \bibinfo{year}{2016}\natexlab{}.
\newblock \showarticletitle{Truncated variance reduction: A unified approach to Bayesian optimization and level-set estimation}. In \bibinfo{booktitle}{\emph{Advances in Neural Information Processing Systems (NeurIPS)}}.
\newblock


\bibitem[Chakareski and Khan(2024)]%
        {ChakareskiK:23}
\bibfield{author}{\bibinfo{person}{J. Chakareski} {and} \bibinfo{person}{M. Khan}.} \bibinfo{year}{2024}\natexlab{}.
\newblock \showarticletitle{Live 360$^\circ$ Video Streaming to Heterogeneous Clients in {5G} Networks}.
\newblock \bibinfo{journal}{\emph{IEEE Trans. Multimedia}}  \bibinfo{volume}{26} (\bibinfo{date}{Aug.} \bibinfo{year}{2024}), \bibinfo{pages}{8860--8873}.
\newblock


\bibitem[Chakareski et~al\mbox{.}(2023)]%
        {ChakareskiKRB:21}
\bibfield{author}{\bibinfo{person}{J. Chakareski}, \bibinfo{person}{M. Khan}, \bibinfo{person}{T. Ropitault}, {and} \bibinfo{person}{S. Blandino}.} \bibinfo{year}{2023}\natexlab{}.
\newblock \showarticletitle{Millimeter Wave and Free-Space-Optics for Future Dual-Connectivity {6DOF} Mobile Multi-User {VR} Streaming}.
\newblock \bibinfo{journal}{\emph{ACM Trans. Multimedia Computing Communications and Applications}} \bibinfo{volume}{19}, \bibinfo{number}{2} (\bibinfo{date}{Feb.} \bibinfo{year}{2023}), \bibinfo{pages}{57:1--25}.
\newblock


\bibitem[Chao et~al\mbox{.}(2020)]%
        {9355679}
\bibfield{author}{\bibinfo{person}{Mengyuan Chao}, \bibinfo{person}{Radu Stoleru}, \bibinfo{person}{Liuyi Jin}, \bibinfo{person}{Shuochao Yao}, \bibinfo{person}{Maxwell Maurice}, {and} \bibinfo{person}{Roger Blalock}.} \bibinfo{year}{2020}\natexlab{}.
\newblock \showarticletitle{AMVP: Adaptive CNN-based Multitask Video Processing on Mobile Stream Processing Platforms}. In \bibinfo{booktitle}{\emph{2020 IEEE/ACM Symposium on Edge Computing (SEC)}}. \bibinfo{pages}{96--109}.
\newblock
\href{https://doi.org/10.1109/SEC50012.2020.00015}{doi:\nolinkurl{10.1109/SEC50012.2020.00015}}


\bibitem[Chowdhury and Gopalan(2017)]%
        {chowdhury2017kernelized}
\bibfield{author}{\bibinfo{person}{Sayak~Ray Chowdhury} {and} \bibinfo{person}{Aditya Gopalan}.} \bibinfo{year}{2017}\natexlab{}.
\newblock \showarticletitle{On kernelized multi-armed bandits}. In \bibinfo{booktitle}{\emph{Proceedings of the 34th International Conference on Machine Learning - Volume 70}} (Sydney, NSW, Australia) \emph{(\bibinfo{series}{ICML'17})}. \bibinfo{publisher}{JMLR.org}, \bibinfo{pages}{844–853}.
\newblock


\bibitem[Dreifuerst et~al\mbox{.}(2021)]%
        {9414155}
\bibfield{author}{\bibinfo{person}{Ryan~M. Dreifuerst}, \bibinfo{person}{Samuel Daulton}, \bibinfo{person}{Yuchen Qian}, \bibinfo{person}{Paul Varkey}, \bibinfo{person}{Maximilian Balandat}, \bibinfo{person}{Sanjay Kasturia}, \bibinfo{person}{Anoop Tomar}, \bibinfo{person}{Ali Yazdan}, \bibinfo{person}{Vish Ponnampalam}, {and} \bibinfo{person}{Robert~W. Heath}.} \bibinfo{year}{2021}\natexlab{}.
\newblock \showarticletitle{Optimizing Coverage and Capacity in Cellular Networks using Machine Learning}. In \bibinfo{booktitle}{\emph{ICASSP 2021 - 2021 IEEE International Conference on Acoustics, Speech and Signal Processing (ICASSP)}}. \bibinfo{pages}{8138--8142}.
\newblock
\href{https://doi.org/10.1109/ICASSP39728.2021.9414155}{doi:\nolinkurl{10.1109/ICASSP39728.2021.9414155}}


\bibitem[Eshratifar et~al\mbox{.}(2021)]%
        {8871124}
\bibfield{author}{\bibinfo{person}{Amir~Erfan Eshratifar}, \bibinfo{person}{Mohammad~Saeed Abrishami}, {and} \bibinfo{person}{Massoud Pedram}.} \bibinfo{year}{2021}\natexlab{}.
\newblock \showarticletitle{JointDNN: An Efficient Training and Inference Engine for Intelligent Mobile Cloud Computing Services}.
\newblock \bibinfo{journal}{\emph{IEEE Transactions on Mobile Computing}} \bibinfo{volume}{20}, \bibinfo{number}{2} (\bibinfo{year}{2021}), \bibinfo{pages}{565--576}.
\newblock
\href{https://doi.org/10.1109/TMC.2019.2947893}{doi:\nolinkurl{10.1109/TMC.2019.2947893}}


\bibitem[Frazier(2018)]%
        {frazier2018tutorialbayesianoptimization}
\bibfield{author}{\bibinfo{person}{Peter~I. Frazier}.} \bibinfo{year}{2018}\natexlab{}.
\newblock \bibinfo{title}{A Tutorial on Bayesian Optimization}.
\newblock
\showeprint[arxiv]{1807.02811}~[stat.ML]
\urldef\tempurl%
\url{https://arxiv.org/abs/1807.02811}
\showURL{%
\tempurl}


\bibitem[Gardner et~al\mbox{.}(2014)]%
        {gardner2014bayesian}
\bibfield{author}{\bibinfo{person}{Jacob Gardner}, \bibinfo{person}{Matt Kusner}, \bibinfo{person}{Zhixiang Xu}, \bibinfo{person}{Kilian Weinberger}, {and} \bibinfo{person}{John Cunningham}.} \bibinfo{year}{2014}\natexlab{}.
\newblock \showarticletitle{Bayesian optimization with inequality constraints}. In \bibinfo{booktitle}{\emph{Proceedings of the 31st International Conference on Machine Learning (ICML)}}.
\newblock


\bibitem[Ghazikor et~al\mbox{.}(2024)]%
        {ghazikor2024channel}
\bibfield{author}{\bibinfo{person}{Masoud Ghazikor}, \bibinfo{person}{Keenan Roach}, \bibinfo{person}{Kenny Cheung}, {and} \bibinfo{person}{Morteza Hashemi}.} \bibinfo{year}{2024}\natexlab{}.
\newblock \showarticletitle{Channel-aware distributed transmission control and video streaming in UAV networks}.
\newblock \bibinfo{journal}{\emph{arXiv preprint arXiv:2408.01885}} (\bibinfo{year}{2024}).
\newblock


\bibitem[Gupta et~al\mbox{.}(2023)]%
        {GuptaCP:20}
\bibfield{author}{\bibinfo{person}{S. Gupta}, \bibinfo{person}{J. Chakareski}, {and} \bibinfo{person}{P. Popovski}.} \bibinfo{year}{2023}\natexlab{}.
\newblock \showarticletitle{{mmWave} Networking and Edge Computing for Scalable 360-Degree Video Multi-User Virtual Reality}.
\newblock \bibinfo{journal}{\emph{IEEE Trans. Image Processing}}  \bibinfo{volume}{32} (\bibinfo{year}{2023}), \bibinfo{pages}{377--391}.
\newblock


\bibitem[Hansen and Ostermeier(2001)]%
        {hansen2001completely}
\bibfield{author}{\bibinfo{person}{Nikolaus Hansen} {and} \bibinfo{person}{Andreas Ostermeier}.} \bibinfo{year}{2001}\natexlab{}.
\newblock \showarticletitle{Completely derandomized self-adaptation in evolution strategies}.
\newblock \bibinfo{journal}{\emph{Evolutionary computation}} \bibinfo{volume}{9}, \bibinfo{number}{2} (\bibinfo{year}{2001}), \bibinfo{pages}{159--195}.
\newblock


\bibitem[Islam et~al\mbox{.}(2021)]%
        {9507488}
\bibfield{author}{\bibinfo{person}{Johirul Islam}, \bibinfo{person}{Tanesh Kumar}, \bibinfo{person}{Ivana Kovacevic}, {and} \bibinfo{person}{Erkki Harjula}.} \bibinfo{year}{2021}\natexlab{}.
\newblock \showarticletitle{Resource-Aware Dynamic Service Deployment for Local IoT Edge Computing: Healthcare Use Case}.
\newblock \bibinfo{journal}{\emph{IEEE Access}}  \bibinfo{volume}{9} (\bibinfo{year}{2021}), \bibinfo{pages}{115868--115884}.
\newblock
\href{https://doi.org/10.1109/ACCESS.2021.3102867}{doi:\nolinkurl{10.1109/ACCESS.2021.3102867}}


\bibitem[Itahara et~al\mbox{.}(2021)]%
        {9685179}
\bibfield{author}{\bibinfo{person}{Sohei Itahara}, \bibinfo{person}{Takayuki Nishio}, {and} \bibinfo{person}{Koji Yamamoto}.} \bibinfo{year}{2021}\natexlab{}.
\newblock \showarticletitle{Packet-Loss-Tolerant Split Inference for Delay-Sensitive Deep Learning in Lossy Wireless Networks}. In \bibinfo{booktitle}{\emph{2021 IEEE Global Communications Conference (GLOBECOM)}}. \bibinfo{pages}{1--6}.
\newblock
\href{https://doi.org/10.1109/GLOBECOM46510.2021.9685179}{doi:\nolinkurl{10.1109/GLOBECOM46510.2021.9685179}}


\bibitem[Jain et~al\mbox{.}(2020)]%
        {mmobile}
\bibfield{author}{\bibinfo{person}{Ish~Kumar Jain}, \bibinfo{person}{Raghav Subbaraman}, \bibinfo{person}{Tejas~Harekrishna Sadarahalli}, \bibinfo{person}{Xiangwei Shao}, \bibinfo{person}{Hou-Wei Lin}, {and} \bibinfo{person}{Dinesh Bharadia}.} \bibinfo{year}{2020}\natexlab{}.
\newblock \showarticletitle{mMobile: Building a mmWave Testbed to Evaluate and Address Mobility Effects}. In \bibinfo{booktitle}{\emph{Proceedings of the 4th ACM Workshop on Millimeter-Wave Networks and Sensing Systems}} (London, United Kingdom) \emph{(\bibinfo{series}{mmNets '20})}. \bibinfo{publisher}{Association for Computing Machinery}, \bibinfo{address}{New York, NY, USA}, Article \bibinfo{articleno}{4}, \bibinfo{numpages}{6}~pages.
\newblock
\showISBNx{9781450380973}
\href{https://doi.org/10.1145/3412060.3418433}{doi:\nolinkurl{10.1145/3412060.3418433}}


\bibitem[Jones et~al\mbox{.}(1993)]%
        {jones1993lipschitzian}
\bibfield{author}{\bibinfo{person}{Donald~R Jones}, \bibinfo{person}{Cary~D Perttunen}, {and} \bibinfo{person}{Bruce~E Stuckman}.} \bibinfo{year}{1993}\natexlab{}.
\newblock \showarticletitle{Lipschitzian optimization without the Lipschitz constant}.
\newblock \bibinfo{journal}{\emph{Journal of optimization Theory and Applications}}  \bibinfo{volume}{79} (\bibinfo{year}{1993}), \bibinfo{pages}{157--181}.
\newblock


\bibitem[Kaggle(2019)]%
        {kaggle_imagenet_mini}
\bibfield{author}{\bibinfo{person}{Kaggle}.} \bibinfo{year}{2019}\natexlab{}.
\newblock \bibinfo{title}{{ImageNet 1000}}.
\newblock \bibinfo{howpublished}{\url{https://www.kaggle.com/datasets/ifigotin/imagenetmini-1000}}.
\newblock


\bibitem[Labriji et~al\mbox{.}(2023)]%
        {10279444}
\bibfield{author}{\bibinfo{person}{Ibtissam Labriji}, \bibinfo{person}{Mattia Merluzzi}, \bibinfo{person}{Fatima~Ezzahra Airod}, {and} \bibinfo{person}{Emilio~Calvanese Strinati}.} \bibinfo{year}{2023}\natexlab{}.
\newblock \showarticletitle{Energy-Efficient Cooperative Inference Via Adaptive Deep Neural Network Splitting at the Edge}. In \bibinfo{booktitle}{\emph{ICC 2023 - IEEE International Conference on Communications}}. \bibinfo{pages}{1712--1717}.
\newblock
\href{https://doi.org/10.1109/ICC45041.2023.10279444}{doi:\nolinkurl{10.1109/ICC45041.2023.10279444}}


\bibitem[Lee et~al\mbox{.}(2023)]%
        {10107635}
\bibfield{author}{\bibinfo{person}{Jaeduk Lee}, \bibinfo{person}{Hojung Lee}, {and} \bibinfo{person}{Wan Choi}.} \bibinfo{year}{2023}\natexlab{}.
\newblock \showarticletitle{Wireless Channel Adaptive DNN Split Inference for Resource-Constrained Edge Devices}.
\newblock \bibinfo{journal}{\emph{IEEE Communications Letters}} \bibinfo{volume}{27}, \bibinfo{number}{6} (\bibinfo{year}{2023}), \bibinfo{pages}{1520--1524}.
\newblock
\href{https://doi.org/10.1109/LCOMM.2023.3269769}{doi:\nolinkurl{10.1109/LCOMM.2023.3269769}}


\bibitem[Li and Bi(2024)]%
        {10478867}
\bibfield{author}{\bibinfo{person}{Xian Li} {and} \bibinfo{person}{Suzhi Bi}.} \bibinfo{year}{2024}\natexlab{}.
\newblock \showarticletitle{Optimal AI Model Splitting and Resource Allocation for Device-Edge Co-Inference in Multi-User Wireless Sensing Systems}.
\newblock \bibinfo{journal}{\emph{IEEE Transactions on Wireless Communications}} \bibinfo{volume}{23}, \bibinfo{number}{9} (\bibinfo{year}{2024}), \bibinfo{pages}{11094--11108}.
\newblock
\href{https://doi.org/10.1109/TWC.2024.3378418}{doi:\nolinkurl{10.1109/TWC.2024.3378418}}


\bibitem[Li et~al\mbox{.}(2024)]%
        {10620728}
\bibfield{author}{\bibinfo{person}{Zuguang Li}, \bibinfo{person}{Wen Wu}, \bibinfo{person}{Shaohua Wu}, {and} \bibinfo{person}{Wei Wang}.} \bibinfo{year}{2024}\natexlab{}.
\newblock \showarticletitle{Adaptive Split Learning over Energy-Constrained Wireless Edge Networks}. In \bibinfo{booktitle}{\emph{IEEE INFOCOM 2024 - IEEE Conference on Computer Communications Workshops (INFOCOM WKSHPS)}}. \bibinfo{pages}{1--6}.
\newblock
\href{https://doi.org/10.1109/INFOCOMWKSHPS61880.2024.10620728}{doi:\nolinkurl{10.1109/INFOCOMWKSHPS61880.2024.10620728}}


\bibitem[Lin et~al\mbox{.}(2024a)]%
        {10529950}
\bibfield{author}{\bibinfo{person}{Zheng Lin}, \bibinfo{person}{Guanqiao Qu}, \bibinfo{person}{Xianhao Chen}, {and} \bibinfo{person}{Kaibin Huang}.} \bibinfo{year}{2024}\natexlab{a}.
\newblock \showarticletitle{Split Learning in 6G Edge Networks}.
\newblock \bibinfo{journal}{\emph{IEEE Wireless Communications}} \bibinfo{volume}{31}, \bibinfo{number}{4} (\bibinfo{year}{2024}), \bibinfo{pages}{170--176}.
\newblock
\href{https://doi.org/10.1109/MWC.014.2300319}{doi:\nolinkurl{10.1109/MWC.014.2300319}}


\bibitem[Lin et~al\mbox{.}(2024b)]%
        {lin2024adaptsfladaptivesplitfederated}
\bibfield{author}{\bibinfo{person}{Zheng Lin}, \bibinfo{person}{Guanqiao Qu}, \bibinfo{person}{Wei Wei}, \bibinfo{person}{Xianhao Chen}, {and} \bibinfo{person}{Kin~K. Leung}.} \bibinfo{year}{2024}\natexlab{b}.
\newblock \bibinfo{title}{AdaptSFL: Adaptive Split Federated Learning in Resource-constrained Edge Networks}.
\newblock
\showeprint[arxiv]{2403.13101}~[cs.LG]
\urldef\tempurl%
\url{https://arxiv.org/abs/2403.13101}
\showURL{%
\tempurl}


\bibitem[Liu et~al\mbox{.}(2024a)]%
        {10646420}
\bibfield{author}{\bibinfo{person}{Hao Liu}, \bibinfo{person}{Mohammed~E. Fouda}, \bibinfo{person}{Ahmed~M. Eltawil}, {and} \bibinfo{person}{Suhaib~A. Fahmy}.} \bibinfo{year}{2024}\natexlab{a}.
\newblock \showarticletitle{Split DNN Inference for Exploiting Near-Edge Accelerators}. In \bibinfo{booktitle}{\emph{2024 IEEE International Conference on Edge Computing and Communications (EDGE)}}. \bibinfo{pages}{84--91}.
\newblock
\href{https://doi.org/10.1109/EDGE62653.2024.00020}{doi:\nolinkurl{10.1109/EDGE62653.2024.00020}}


\bibitem[Liu et~al\mbox{.}(2024b)]%
        {9933908}
\bibfield{author}{\bibinfo{person}{Zhicheng Liu}, \bibinfo{person}{Jinduo Song}, \bibinfo{person}{Chao Qiu}, \bibinfo{person}{Xiaofei Wang}, \bibinfo{person}{Xu Chen}, \bibinfo{person}{Qiang He}, {and} \bibinfo{person}{Hao Sheng}.} \bibinfo{year}{2024}\natexlab{b}.
\newblock \showarticletitle{Hastening Stream Offloading of Inference via Multi-Exit DNNs in Mobile Edge Computing}.
\newblock \bibinfo{journal}{\emph{IEEE Transactions on Mobile Computing}} \bibinfo{volume}{23}, \bibinfo{number}{1} (\bibinfo{year}{2024}), \bibinfo{pages}{535--548}.
\newblock
\href{https://doi.org/10.1109/TMC.2022.3218724}{doi:\nolinkurl{10.1109/TMC.2022.3218724}}


\bibitem[Lu et~al\mbox{.}(2022)]%
        {lu2022surrogatemodelingbayesianoptimization}
\bibfield{author}{\bibinfo{person}{Qin Lu}, \bibinfo{person}{Konstantinos~D. Polyzos}, \bibinfo{person}{Bingcong Li}, {and} \bibinfo{person}{Georgios~B. Giannakis}.} \bibinfo{year}{2022}\natexlab{}.
\newblock \bibinfo{title}{Surrogate modeling for Bayesian optimization beyond a single Gaussian process}.
\newblock
\showeprint[arxiv]{2205.14090}~[stat.ML]
\urldef\tempurl%
\url{https://arxiv.org/abs/2205.14090}
\showURL{%
\tempurl}


\bibitem[Maggi et~al\mbox{.}(2021)]%
        {9430561}
\bibfield{author}{\bibinfo{person}{Lorenzo Maggi}, \bibinfo{person}{Alvaro Valcarce}, {and} \bibinfo{person}{Jakob Hoydis}.} \bibinfo{year}{2021}\natexlab{}.
\newblock \showarticletitle{Bayesian Optimization for Radio Resource Management: Open Loop Power Control}.
\newblock \bibinfo{journal}{\emph{IEEE Journal on Selected Areas in Communications}} \bibinfo{volume}{39}, \bibinfo{number}{7} (\bibinfo{year}{2021}), \bibinfo{pages}{1858--1871}.
\newblock
\href{https://doi.org/10.1109/JSAC.2021.3078490}{doi:\nolinkurl{10.1109/JSAC.2021.3078490}}


\bibitem[Mallick et~al\mbox{.}(2011)]%
        {5963233}
\bibfield{author}{\bibinfo{person}{Shankhanaad Mallick}, \bibinfo{person}{Kundan Kandhway}, \bibinfo{person}{Mohammad~M. Rashid}, {and} \bibinfo{person}{Vijay~K. Bhargava}.} \bibinfo{year}{2011}\natexlab{}.
\newblock \showarticletitle{Power Allocation for Decode-and-Forward Cellular Relay Network with Channel Uncertainty}. In \bibinfo{booktitle}{\emph{2011 IEEE International Conference on Communications (ICC)}}. \bibinfo{pages}{1--5}.
\newblock
\href{https://doi.org/10.1109/icc.2011.5963233}{doi:\nolinkurl{10.1109/icc.2011.5963233}}


\bibitem[Mao et~al\mbox{.}(2022)]%
        {9224971}
\bibfield{author}{\bibinfo{person}{Sun Mao}, \bibinfo{person}{Jinsong Wu}, \bibinfo{person}{Lei Liu}, \bibinfo{person}{Dapeng Lan}, {and} \bibinfo{person}{Amir Taherkordi}.} \bibinfo{year}{2022}\natexlab{}.
\newblock \showarticletitle{Energy-Efficient Cooperative Communication and Computation for Wireless Powered Mobile-Edge Computing}.
\newblock \bibinfo{journal}{\emph{IEEE Systems Journal}} \bibinfo{volume}{16}, \bibinfo{number}{1} (\bibinfo{year}{2022}), \bibinfo{pages}{287--298}.
\newblock
\href{https://doi.org/10.1109/JSYST.2020.3020474}{doi:\nolinkurl{10.1109/JSYST.2020.3020474}}


\bibitem[Mao et~al\mbox{.}(2017)]%
        {mao2017surveymobileedgecomputing}
\bibfield{author}{\bibinfo{person}{Yuyi Mao}, \bibinfo{person}{Changsheng You}, \bibinfo{person}{Jun Zhang}, \bibinfo{person}{Kaibin Huang}, {and} \bibinfo{person}{Khaled~B. Letaief}.} \bibinfo{year}{2017}\natexlab{}.
\newblock \bibinfo{title}{A Survey on Mobile Edge Computing: The Communication Perspective}.
\newblock
\showeprint[arxiv]{1701.01090}~[cs.IT]
\urldef\tempurl%
\url{https://arxiv.org/abs/1701.01090}
\showURL{%
\tempurl}


\bibitem[Miao et~al\mbox{.}(2020)]%
        {9359147}
\bibfield{author}{\bibinfo{person}{Weiwei Miao}, \bibinfo{person}{Zeng Zeng}, \bibinfo{person}{Lei Wei}, \bibinfo{person}{Shihao Li}, \bibinfo{person}{Chengling Jiang}, {and} \bibinfo{person}{Zhen Zhang}.} \bibinfo{year}{2020}\natexlab{}.
\newblock \showarticletitle{Adaptive DNN Partition in Edge Computing Environments}. In \bibinfo{booktitle}{\emph{2020 IEEE 26th International Conference on Parallel and Distributed Systems (ICPADS)}}. \bibinfo{pages}{685--690}.
\newblock
\href{https://doi.org/10.1109/ICPADS51040.2020.00097}{doi:\nolinkurl{10.1109/ICPADS51040.2020.00097}}


\bibitem[Mudvari et~al\mbox{.}(2024)]%
        {mudvari2024adaptive}
\bibfield{author}{\bibinfo{person}{Akrit Mudvari}, \bibinfo{person}{Antero Vainio}, \bibinfo{person}{Iason Ofeidis}, \bibinfo{person}{Sasu Tarkoma}, {and} \bibinfo{person}{Leandros Tassiulas}.} \bibinfo{year}{2024}\natexlab{}.
\newblock \showarticletitle{Adaptive compression-aware split learning and inference for enhanced network efficiency}.
\newblock \bibinfo{journal}{\emph{ACM Transactions on Internet Technology}} \bibinfo{volume}{24}, \bibinfo{number}{4} (\bibinfo{year}{2024}), \bibinfo{pages}{1--26}.
\newblock


\bibitem[Ni et~al\mbox{.}(2025)]%
        {ni2025pfedwn}
\bibfield{author}{\bibinfo{person}{Zhou Ni}, \bibinfo{person}{Masoud Ghazikor}, {and} \bibinfo{person}{Morteza Hashemi}.} \bibinfo{year}{2025}\natexlab{}.
\newblock \showarticletitle{pFedWN: A Personalized Federated Learning Framework for D2D Wireless Networks with Heterogeneous Data}.
\newblock \bibinfo{journal}{\emph{arXiv preprint arXiv:2501.09822}} (\bibinfo{year}{2025}).
\newblock


\bibitem[Papenmeier et~al\mbox{.}(2024)]%
        {papenmeier2024bouncereliablehighdimensionalbayesian}
\bibfield{author}{\bibinfo{person}{Leonard Papenmeier}, \bibinfo{person}{Luigi Nardi}, {and} \bibinfo{person}{Matthias Poloczek}.} \bibinfo{year}{2024}\natexlab{}.
\newblock \bibinfo{title}{Bounce: Reliable High-Dimensional Bayesian Optimization for Combinatorial and Mixed Spaces}.
\newblock
\showeprint[arxiv]{2307.00618}~[cs.LG]
\urldef\tempurl%
\url{https://arxiv.org/abs/2307.00618}
\showURL{%
\tempurl}


\bibitem[Safaeipour and Hashemi(2024)]%
        {sac}
\bibfield{author}{\bibinfo{person}{Fatemeh~Zahra Safaeipour} {and} \bibinfo{person}{Morteza Hashemi}.} \bibinfo{year}{2024}\natexlab{}.
\newblock \bibinfo{title}{Semantic-Aware and Goal-Oriented Communications for Object Detection in Wireless End-to-End Image Transmission}.
\newblock
\showeprint[arxiv]{2402.01064}~[cs.IT]
\urldef\tempurl%
\url{https://arxiv.org/abs/2402.01064}
\showURL{%
\tempurl}


\bibitem[Simonyan and Zisserman(2015)]%
        {vgg}
\bibfield{author}{\bibinfo{person}{Karen Simonyan} {and} \bibinfo{person}{Andrew Zisserman}.} \bibinfo{year}{2015}\natexlab{}.
\newblock \bibinfo{title}{Very Deep Convolutional Networks for Large-Scale Image Recognition}.
\newblock
\showeprint[arxiv]{1409.1556}~[cs.CV]
\urldef\tempurl%
\url{https://arxiv.org/abs/1409.1556}
\showURL{%
\tempurl}


\bibitem[Singhal et~al\mbox{.}(2024)]%
        {10621218}
\bibfield{author}{\bibinfo{person}{C. Singhal}, \bibinfo{person}{Y. Wu}, \bibinfo{person}{F. Malandrino}, \bibinfo{person}{M. Levorato}, {and} \bibinfo{person}{C.~F. Chiasserini}.} \bibinfo{year}{2024}\natexlab{}.
\newblock \showarticletitle{Resource-aware Deployment of Dynamic DNNs over Multi-tiered Interconnected Systems}. In \bibinfo{booktitle}{\emph{IEEE INFOCOM 2024 - IEEE Conference on Computer Communications}}. \bibinfo{pages}{1621--1630}.
\newblock
\href{https://doi.org/10.1109/INFOCOM52122.2024.10621218}{doi:\nolinkurl{10.1109/INFOCOM52122.2024.10621218}}


\bibitem[Srinivas et~al\mbox{.}(2010)]%
        {srinivas2010gaussian}
\bibfield{author}{\bibinfo{person}{Niranjan Srinivas}, \bibinfo{person}{Andreas Krause}, \bibinfo{person}{Sham Kakade}, {and} \bibinfo{person}{Matthias Seeger}.} \bibinfo{year}{2010}\natexlab{}.
\newblock \showarticletitle{Gaussian process optimization in the bandit setting: no regret and experimental design}. In \bibinfo{booktitle}{\emph{Proceedings of the 27th International Conference on International Conference on Machine Learning}} (Haifa, Israel) \emph{(\bibinfo{series}{ICML'10})}. \bibinfo{publisher}{Omnipress}, \bibinfo{pages}{1015–1022}.
\newblock
\showISBNx{9781605589077}


\bibitem[Srinivas et~al\mbox{.}(2012)]%
        {Srinivas_2012}
\bibfield{author}{\bibinfo{person}{Niranjan Srinivas}, \bibinfo{person}{Andreas Krause}, \bibinfo{person}{Sham~M. Kakade}, {and} \bibinfo{person}{Matthias~W. Seeger}.} \bibinfo{year}{2012}\natexlab{}.
\newblock \showarticletitle{Information-Theoretic Regret Bounds for Gaussian Process Optimization in the Bandit Setting}.
\newblock \bibinfo{journal}{\emph{IEEE Transactions on Information Theory}} \bibinfo{volume}{58}, \bibinfo{number}{5} (\bibinfo{date}{May} \bibinfo{year}{2012}), \bibinfo{pages}{3250–3265}.
\newblock
\showISSN{1557-9654}
\href{https://doi.org/10.1109/tit.2011.2182033}{doi:\nolinkurl{10.1109/tit.2011.2182033}}


\bibitem[Thapa et~al\mbox{.}(2022)]%
        {thapa2020splitfed}
\bibfield{author}{\bibinfo{person}{Chandra Thapa}, \bibinfo{person}{M.~A.~P. Chamikara}, \bibinfo{person}{Seyit Camtepe}, {and} \bibinfo{person}{Lichao Sun}.} \bibinfo{year}{2022}\natexlab{}.
\newblock \showarticletitle{SplitFed: When Federated Learning Meets Split Learning}.
\newblock  (\bibinfo{year}{2022}).
\newblock
\showeprint[arxiv]{2004.12088}~[cs.LG]
\urldef\tempurl%
\url{https://arxiv.org/abs/2004.12088}
\showURL{%
\tempurl}


\bibitem[Wang et~al\mbox{.}(2018)]%
        {8234686}
\bibfield{author}{\bibinfo{person}{Feng Wang}, \bibinfo{person}{Jie Xu}, \bibinfo{person}{Xin Wang}, {and} \bibinfo{person}{Shuguang Cui}.} \bibinfo{year}{2018}\natexlab{}.
\newblock \showarticletitle{Joint Offloading and Computing Optimization in Wireless Powered Mobile-Edge Computing Systems}.
\newblock \bibinfo{journal}{\emph{IEEE Transactions on Wireless Communications}} \bibinfo{volume}{17}, \bibinfo{number}{3} (\bibinfo{year}{2018}), \bibinfo{pages}{1784--1797}.
\newblock
\href{https://doi.org/10.1109/TWC.2017.2785305}{doi:\nolinkurl{10.1109/TWC.2017.2785305}}


\bibitem[Wu et~al\mbox{.}(2023)]%
        {10040976}
\bibfield{author}{\bibinfo{person}{Wen Wu}, \bibinfo{person}{Mushu Li}, \bibinfo{person}{Kaige Qu}, \bibinfo{person}{Conghao Zhou}, \bibinfo{person}{Xuemin Shen}, \bibinfo{person}{Weihua Zhuang}, \bibinfo{person}{Xu Li}, {and} \bibinfo{person}{Weisen Shi}.} \bibinfo{year}{2023}\natexlab{}.
\newblock \showarticletitle{Split Learning Over Wireless Networks: Parallel Design and Resource Management}.
\newblock \bibinfo{journal}{\emph{IEEE Journal on Selected Areas in Communications}} \bibinfo{volume}{41}, \bibinfo{number}{4} (\bibinfo{year}{2023}), \bibinfo{pages}{1051--1066}.
\newblock
\href{https://doi.org/10.1109/JSAC.2023.3242704}{doi:\nolinkurl{10.1109/JSAC.2023.3242704}}


\bibitem[Yan et~al\mbox{.}(2024)]%
        {10234224}
\bibfield{author}{\bibinfo{person}{Jia Yan}, \bibinfo{person}{Qin Lu}, {and} \bibinfo{person}{Georgios~B. Giannakis}.} \bibinfo{year}{2024}\natexlab{}.
\newblock \showarticletitle{Bayesian Optimization for Online Management in Dynamic Mobile Edge Computing}.
\newblock \bibinfo{journal}{\emph{IEEE Transactions on Wireless Communications}} \bibinfo{volume}{23}, \bibinfo{number}{4} (\bibinfo{year}{2024}), \bibinfo{pages}{3425--3436}.
\newblock
\href{https://doi.org/10.1109/TWC.2023.3307875}{doi:\nolinkurl{10.1109/TWC.2023.3307875}}


\bibitem[Yang et~al\mbox{.}(2023)]%
        {10038613}
\bibfield{author}{\bibinfo{person}{Yuzhi Yang}, \bibinfo{person}{Zhaoyang Zhang}, \bibinfo{person}{Yuqing Tian}, \bibinfo{person}{Zhaohui Yang}, \bibinfo{person}{Chongwen Huang}, \bibinfo{person}{Caijun Zhong}, {and} \bibinfo{person}{Kai-Kit Wong}.} \bibinfo{year}{2023}\natexlab{}.
\newblock \showarticletitle{Over-the-Air Split Machine Learning in Wireless MIMO Networks}.
\newblock \bibinfo{journal}{\emph{IEEE Journal on Selected Areas in Communications}} \bibinfo{volume}{41}, \bibinfo{number}{4} (\bibinfo{year}{2023}), \bibinfo{pages}{1007--1022}.
\newblock
\href{https://doi.org/10.1109/JSAC.2023.3242701}{doi:\nolinkurl{10.1109/JSAC.2023.3242701}}


\bibitem[Yao et~al\mbox{.}(2025)]%
        {11016266}
\bibfield{author}{\bibinfo{person}{Jiacheng Yao}, \bibinfo{person}{Wei Xu}, \bibinfo{person}{Guangxu Zhu}, \bibinfo{person}{Kaibin Huang}, {and} \bibinfo{person}{Shuguang Cui}.} \bibinfo{year}{2025}\natexlab{}.
\newblock \showarticletitle{Energy-Efficient Edge Inference in Integrated Sensing, Communication, and Computation Networks}.
\newblock \bibinfo{journal}{\emph{IEEE Journal on Selected Areas in Communications}} (\bibinfo{year}{2025}), \bibinfo{pages}{1--1}.
\newblock
\href{https://doi.org/10.1109/JSAC.2025.3574612}{doi:\nolinkurl{10.1109/JSAC.2025.3574612}}


\bibitem[Ying(2017)]%
        {7890237}
\bibfield{author}{\bibinfo{person}{Beihua Ying}.} \bibinfo{year}{2017}\natexlab{}.
\newblock \showarticletitle{An adaptive compression algorithm for energy-efficient wireless sensor networks}. In \bibinfo{booktitle}{\emph{2017 19th International Conference on Advanced Communication Technology (ICACT)}}. \bibinfo{pages}{861--868}.
\newblock
\href{https://doi.org/10.23919/ICACT.2017.7890237}{doi:\nolinkurl{10.23919/ICACT.2017.7890237}}


\bibitem[Zhang et~al\mbox{.}(2024)]%
        {zhang2024proximal}
\bibfield{author}{\bibinfo{person}{Chen Zhang}, \bibinfo{person}{Celimuge Wu}, \bibinfo{person}{Min Lin}, \bibinfo{person}{Yangfei Lin}, {and} \bibinfo{person}{William Liu}.} \bibinfo{year}{2024}\natexlab{}.
\newblock \showarticletitle{Proximal Policy Optimization for Efficient D2D-Assisted Computation Offloading and Resource Allocation in Multi-Access Edge Computing}.
\newblock \bibinfo{journal}{\emph{Future Internet}} \bibinfo{volume}{16}, \bibinfo{number}{1} (\bibinfo{year}{2024}), \bibinfo{pages}{19}.
\newblock


\bibitem[Zhang et~al\mbox{.}(2021)]%
        {9384272}
\bibfield{author}{\bibinfo{person}{Weiting Zhang}, \bibinfo{person}{Dong Yang}, \bibinfo{person}{Haixia Peng}, \bibinfo{person}{Wen Wu}, \bibinfo{person}{Wei Quan}, \bibinfo{person}{Hongke Zhang}, {and} \bibinfo{person}{Xuemin Shen}.} \bibinfo{year}{2021}\natexlab{}.
\newblock \showarticletitle{Deep Reinforcement Learning Based Resource Management for DNN Inference in Industrial IoT}.
\newblock \bibinfo{journal}{\emph{IEEE Transactions on Vehicular Technology}} \bibinfo{volume}{70}, \bibinfo{number}{8} (\bibinfo{year}{2021}), \bibinfo{pages}{7605--7618}.
\newblock
\href{https://doi.org/10.1109/TVT.2021.3068255}{doi:\nolinkurl{10.1109/TVT.2021.3068255}}


\end{thebibliography}

\end{document}